 \documentclass[final,authoryear,3p,12pt]{elsarticle}

\usepackage{graphicx}
\usepackage{epstopdf}
\usepackage{booktabs}
\usepackage{natbib}
\usepackage{fancyhdr}
\usepackage{amsmath}
\usepackage{amsfonts}
\usepackage{multirow}
\usepackage{lscape}
\usepackage{setspace}
\usepackage{color}
\usepackage{hyperref}
\usepackage{color,soul}
\sethlcolor{green}
\usepackage{ifthen}
\usepackage[latin1]{inputenc}
\usepackage{titletoc}
\usepackage{eso-pic}
\definecolor{123}{rgb}{.9,.9,.9}
\hypersetup{
colorlinks=false,
citecolor=blue,
linkbordercolor={1 1 1}, 
citebordercolor={1 1 1},
urlbordercolor={1 1 1}
}
 
\usepackage{amssymb}
 \usepackage{amsthm}
 \usepackage{xcolor}

\newtheorem{prop}{Proposition}
\def \ZN {\mathbb Z_N}

\begin{document}

\begin{frontmatter}

%
%
\title{Revisiting the fractional cointegrating dynamics of implied-realized volatility relation with wavelet band spectrum regression\tnoteref{label1} }
%
%
\author[ies,utia]{Jozef Barun\'ik\corref{cor2} } \ead{barunik@utia.cas.cz}
\author[ies]{Michaela Barun\'ikov\'a}
\address[ies]{Institute of Economic Studies, Charles University, Opletalova 21, 110 00, Prague,  CR.}
\address[utia]{Institute of Information Theory and Automation, Academy of Sciences of the Czech Republic, Pod Vodarenskou Vezi 4, 182 00, Prague, Czech Republic}
\tnotetext[label1]{We are grateful to Lukas Vacha as well as seminar participants at the Computational and Financial Econometrics in London (December 2011) for many useful discussions. Jozef Barunik gratefully acknowledges the support of the Czech Science Foundation project No. P402/12/G097 DYME - ``Dynamic Models in Economics". Michaela Barunikova greatly acknowledges the support of the Czech Science Foundation project No. 402/09/0732 and support of the GAUK 378711 project.}
\begin{abstract}
This paper revisits the fractional cointegrating relationship between ex-ante implied volatility and ex-post realized volatility. We argue that the concept of corridor implied volatility (CIV) should be used instead of the popular model-free option-implied volatility (MFIV) when assessing the fractional cointegrating relation as the latter may introduce bias to the estimation. For the realized volatility, we use recently proposed methods which are robust to noise as well as jumps and interestingly we find that it does not affect the implied-realized volatility relation. In addition, we develop a new tool for the estimation of fractional cointegrating relation between implied and realized volatility based on wavelets, a wavelet band least squares (WBLS). The main advantage of WBLS in comparison to other frequency domain methods is that it allows us to work conveniently with potentially non-stationary volatility due to the properties of wavelets. We study the dynamics of the relationship in the time-frequency domain with the wavelet coherence confirming that the dependence comes solely from the lower frequencies of the spectra. Motivated by this result we estimate the relationship only on this part of the spectra using WBLS and compare our results to the fully modified narrow-band least squares (FMNBLS) based on the Fourier frequencies. In the estimation, we use the S\&P 500 and DAX monthly and bi-weekly option prices covering the recent financial crisis and we conclude that in the long-run, volatility inferred from the option prices using the corridor implied volatility (CIV) provides an unbiased forecast of the realized volatility.
\end{abstract}
\end{frontmatter}

\section{Introduction}

Option prices are widely believed to carry information relating to expectations of market participants about the future movement of the underlying asset prices in financial markets, mostly its volatility. The volatility implied by an option's price is the forecast of the future return volatility over the remaining life of the relevant option if the option markets are efficient. Early papers studying the phenomenon of implied-realized volatility relation use volatility implied by option pricing models -- most commonly \cite{BS73} or \cite{HW87} -- and come to a conclusion that volatility inferred from the option markets is a biased predictor of the stock return volatility \citep{DL92,LL93,CF93,J95}.

In contrast, \cite{christensen1998,christensen2002} first notes that ex-ante implied volatility in fact is an unbiased and efficient forecast of ex-post volatility after the 1987 stock market crash, while they point to large bias before the 1987 crash. Using wide variety of methods authors show that information content of option implied volatility is superior to that of the past volatility and it is less biased (although still biased) predictor of future realized volatility than has been previously shown. Authors shed some light on the dubiety about the informational content of option implied volatility by specifying the sources of errors of previous research, for example choice of particular option contracts to extract volatility from and lower liquidity of option market than in underlying asset market. Moreover, overlapping data errors cause cross-correlation in the volatility series, which stems from overlapping period between the current implied volatility and future implied volatility. Yet, there is a problem with non overlapping samples, as they drastically reduce the data used. Last, but not least, mismatching errors include for example maturity mismatch (options with longer expiration are used to predict day/week ahead realized volatilities). In the light of these methodological issues, \cite{christensen2002} conclude that option implied volatility is more efficient forecast for future realized volatility than historical volatility, but it does not subsume all information contained in historical volatility and it is an upward biased forecast for future realized. Still, the approach to infer the volatility from the option markets is model based, authors use a modification of \cite{BS73}.


Strikingly simple method to extract volatilities from options across all strike prices, model-free implied volatility (MFIV) was introduced by \cite{BN00}. Unlike the traditional concept, model-free implied volatility is not based on any specific option pricing model and it is derived from no-arbitrage conditions.  \cite{Jiang2005} extended the simple measure of implied volatility to all martingale asset price processes and express the formulae in forward rather than spot prices. Most notably, \cite{Jiang2005} first find that the MFIV subsumes all information contained by historical and \cite{BS73} implied volatility and is more efficient forecast of future realized volatility. Informational content of option implied volatility in the subsequent research is analyzed using the model-free measure\footnote{Many studies use the Chicago Board Option Exchange (CBOE) Volatility Index (VIX) as a proxy for model-free implied volatility of S\&P 500. Introduced by the CBOE in 1993, its methodology was revised in 2003 using a new model-free measure of expected volatility thus is can be used conveniently.}.  

Finally, \cite{Andersen2007} and \cite{Andersen2011} argue that MFIV computation brings serious practical limitations yielding an inaccurate measurement tool. The main problem is the lack of liquid options with strike prices covering the entire return distribution including its tails. Authors advocate using a limited strike ranges at a given point in time instead. The concept is called model-free corridor implied volatility index (CIV), previously introduced by \cite{carrmadan1998}. While different measures can be obtained depending on the width and positioning of the strike range, \cite{Andersen2011} advocate fixing the range of strikes at a level that provides broad coverage but avoids excessive extrapolation of noisy or non-existing quotes for far out-of-the-money options.

When assessing the efficiency of implied volatility forecasts, one needs to have a return volatility at hand. However, actual volatility has not been directly observable variable for a long time. In recent years, as a consequence of an increased availability of  high-frequency data, another subject has brought new insight into the implied-realized volatility relation; the concept of realized volatility, which is simply observed sample variance of returns. \cite{abdl2003} and \cite{barndorff2004b} have shown that realized variance provides a consistent nonparametric measure of price variability over a given time interval. Immense literature studying the realized volatility emerged in the past decade discussing the impact of noise as well as jumps in the volatility measurement concluding that realized volatility is unbiased and consistent measure of quadratic variation only in case we assume no market microstructure noise in the process. Literature also argues that it is important to separate jump process and use the estimator robust to noise to recover a true underlying volatility. For this, we use recently proposed jump wavelet two scale realized volatility estimator (JWTSRV) proposed by \cite{barunik}, which compares to other estimators used in the literature very well. JWTSRV is able to estimate the jumps consistently and the estimator is robust to noise. Forecasting power of the estimator is studied using a Realized GARCH framework in \cite{barunik2}.

Using more and more precise measures, recent literature suggests that predictive regression between implied volatility  and realized volatility is an cointegrating relation and OLS estimation should be avoided as it will result in biased estimates \citep{Bandi2006,NielsenFrederiksen}. Using a spectral methods, both studies confirm that in a long-run, implied volatility is an unbiased predictor of realized volatility. Still, the results do not say anything about short-term unbiasedness and they rely on Black-Scholes implied volatility only. Generally,  band spectrum regression may be useful tool in the situation when we believe the relationship between variables is dependent on frequency. The concept was introduced in econometrics by \cite{engle1974} and further shown to be useful for estimation of cointegrating regressions \citep{phillips1991,marinuccirobinson2001}. 

While \cite{Bandi2006,NielsenFrederiksen,Kellard2010} use Fourier transform in order to estimate the relation in the frequency domain, we contribute to the literature by proposing the band regression on the spectrum estimated by wavelet coefficients. The wavelet transform offers localized frequency decomposition, providing information about frequency components. As a result, wavelets have significant advantages over basic Fourier analysis when the object under study is locally stationary and inhomogeneous -- see \cite{Gencay2002, PercivalWalden2000,Ramsay2002}. This can be a crucial property as implied-realized volatility cointegrating relation may potentially lie in a non-stationary region  \citep{Kellard2010}.

Wavelets also allow us to study the relationship in the time-frequency domain. We motivate this dynamics by estimating the wavelet coherence measure to study the implied-realized relation. While wavelet coherence may be used as the ``lens" into the relationship which shows the dynamics through time as well as frequencies at once, a newly proposed wavelet band spectral regression allows us to estimate the relationship. 

The contribution of this paper is twofold. First, we develop a new time-series technique to validate the unbiasedness of ex-ante implied volatility as a predictor of ex-post realized volatility, the wavelet band least squares (WBLS). Second, we emphasize the importance of the implied volatility measure in studying the cointegrating relation. We compare MFIV and recently proposed CIV as measures of option implied volatilities with realized volatility and recently proposed jump-wavelet realized volatility (JWTSRV) capable of separating the continuous part of the volatility from jumps as well as noise. We argue that it is crucial to use a proper measures for finding the answer to the question whether the option implied volatility is an efficient forecast of the realized volatility. The methods are applied on German DAX and U.S. S\&P 500 stock market indices covering the 2008 financial crisis with abrupt changes in prices. Unlike the previous studies, we use both call and put options and we use options with monthly as well as bi-weekly maturities.

The paper is organized as follows: second section describes the volatility measurement. It starts generally with concept of realized volatility and introduces model-free implied volatility (MFIV) as well as corridor implied volatility (CIV) subsequently. After the introduction of the data in the third section, fourth section begins to study the time-frequency dependence of implied and realized volatility using wavelet coherence, introduces our wavelet band least squares (WBLS) to study the fractionally cointegrated series and summarizes narrow band least squares (NBLS) and fully modified narrow band least squares (FMNBLS) methodology. Section five discusses the results while last section concludes. As we use wavelets extensively but do not want to distract the reader from the main text, we also provide the short introduction to the wavelet methodology in the appendix.


\section{Volatility measurement}

Consider a univariate risky logarithmic asset price process $p_t$ defined on a complete probability space $ (\Omega,\mathcal{F},\mathbb{P})$. The price process evolves in continuous time $t$ over the interval $\left[0,T\right]$, where $T$ is a finite positive integer according to a jump diffusion process:
\begin{equation}
\label{eqdpt}
dp_t=\mu_t dt+\sigma_t dW_t+\xi_t dq_t,
\end{equation}
where $\mu_t$ is a predictable mean, $\sigma_t$ strictly positive volatility process, $W_t$ is standard Brownian motion, $\xi_t dq_t$ is a random jump process allowing for occasional jumps in $p_t$ and $q$ is a Poisson process uncorrelated with $W$ and governed by the constant jump intensity $\lambda$. The magnitude of the jump in the return process is controlled by factor $\xi_t \sim N(\bar\xi,\sigma^2_{\xi})$.

Generally, we assume the latent logarithmic asses price process is contamined with microstructure noise. Let  $y_t$  be the observed log prices, which will be equal to the latent, so-called ``true" log-price process $p_t$ in Eq. (\ref{eqdpt}) and will contain microstructure noise $\epsilon_t$, a zero mean $i.i.d.$ noise with variance $\eta^2$
 \begin{equation}
 \label{eqnoise}
 y_t=p_t+\epsilon_t.
 \end{equation}
The main object of interest in financial econometrics is the estimated integrated variance of the latent price process,  $IV_{t,h}=\int_{t-h}^t \sigma_t^2dt$. Quadratic return variation over the $\left[t-h,t\right]$ time interval, $0 \le h \le t \le T$ is 
\begin{equation}
\label{waveqvjump}
QV_{t,h}=\underbrace{\int_{t-h}^t \sigma_s^2 ds}_{\mbox{$IV_{t,h}$}}+ \underbrace{\sum_{t-h \le s \le t} J_s^2}_{\mbox{$JV_{t,h}$}}.
\end{equation}
Thus, quadratic variation $QV_{t,h}$ is equal to the integrated volatility of the continuous path and the sum of jumps with size $J_s$.


\subsection{Realized volatility}

Recently popularized simple measure of quadratic variation -- realized variance -- is consistent and unbiased estimator of the quadratic variation if the sampling goes to infinity \citep{ab98,abdl2003, barndorff2001,barndorff2002a,barndorff2002}. The realized variance over $\left[t-h,t\right]$, for $0\le h \le t\le T$, is defined by
\begin{equation}
\label{rv}
\widehat{RV}_{t,h}=\sum_{i=1}^n r_{t-h+\left(\frac{i}{n}\right)h}^2,
\end{equation}
where $n$ is the number of observations in $\left[t-h,t\right]$. While the realized variance measure is widely used due to its simplicity, it estimates the whole part of the quadratic variation and is subject to the bias from the microstructure noise. In fact when the realized variance is used to measure the volatility (its square root), it will measure volatility of $y_t$ from Eq. (\ref{eqnoise}) together with jumps. Still, the main interest is to measure the integrated variance part $IV_{t,h}$ hence more sophisticated estimators need to be used. 

Literature developed several estimators dealing with microstructure noise and jumps recently. For example \cite{zhang2005} propose the solution to the problem of noise by introducing the two-scale realized volatility (TSRV henceforth) estimator. Another estimator, which is able to deal with the noise is the realized kernels (RK) introduced by \cite{barndorff2008}. \cite{barndorff2004,barndorff2006} developed a very powerful and complete way of detecting the presence of jumps in high-frequency data, bipower variation. The basic idea is to compare two measures of the integrated variance, one containing the jump variation and the other being robust to jumps and hence containing only the integrated variation part. 

More recently, jump adjusted wavelet two scale realized volatility (JWTSRV) has been proposed to measure the integrated variance in the presence of jumps and noise by \cite{barunik}. JWTSRV is able to consistently estimate jumps using wavelet transform and is also robust to microstructure noise thanks to a \cite{zhang2005}'s framework. Let us introduce the estimator.

Starting with the jump detection, \cite{barunik} utilize the methodology proposed by \cite{fanwang2008} who use the wavelet jump detection to the deterministic functions with $i.i.d.$ additive noise $\epsilon_t$ of \cite{wang95}. For the estimation of jump location the universal threshold of \cite{donoho} on the $1^{st}$ level wavelet coefficients of $y_t$ over $[t-h,t]$, $\widetilde{\mathcal{W}}_{1,k}$ is used\footnote{Not to distract the reader from the main text, we provide the necessary introduction to wavelet analysis in an Appendix \ref{appWavelets}.}. If for some $\widetilde{\mathcal{W}}_{1,k}$, $ |\widetilde{\mathcal{W}}_{1,k}| > d \sqrt{2 \log n}$, then $\hat{\tau}_l=\{k\}$ is the estimated jump location with size $\bar{y}_{\hat{\tau}_{l+}}-\bar{y}_{\hat{\tau}_{l-}}$ (averages over $[\hat{\tau}_l,\hat{\tau}_l+\delta_n]$ and $[\hat{\tau}_l,\hat{\tau}_l-\delta_n]$, respectively, with $\delta_n > 0$ being the small neighborhood of the estimated jump location $\hat{\tau}_l \pm \delta_n$) and where $d$ is median absolute deviation estimator defined as $(2^{1/2})median\{|\widetilde{\mathcal{W}}_{1,k}|,k=1,.\dots,n\}/0.6745$, for more details see \cite{PercivalWalden2000}.

Using the result of \cite{fanwang2008}, the jump variation is then estimated by the sum of the squares of all the estimated jump sizes:
 \begin{equation}
 \label{wjwequation}
\widehat{JV}_{t,h}^W=\sum_{l=1}^{N_t} (\bar{y}_{t,h,\hat{\tau}_{l+}}-\bar{y}_{t,h,\hat{\tau}_{l-}})^2,
  \end{equation}
 thus we are able to estimate the jump variation from the process consistently with the convergence rate $N^{-1/4}$. In the following analysis, we will be able to separate the continuous part of the price process containing noise from the jump variation. This result can be found in \cite{fanwang2008} and it states that the jump-adjusted process $y_{t,h}^{(J)}=y_{t,h
}-\widehat{JV}_{t,h}^W$ converges in probability to the continuous part without jumps. Thus, if we are able to deal with the noise in $y_{t,h}^{(J)}$, we will be able to estimate the $IV_{t,h}$.

Following \cite{barunik}, we define the jump-adjusted wavelet two-scale realized variance (JWTSRV) estimator over $\left[t-h,t\right]$, for $0 \le h \le t \le T$, on the observed jump-adjusted data, $y_{t,h}^{(J)}=y_{t,h}-\sum_{l=1}^{N_t} J_l$ as: 

\begin{equation}
\label{jwtsrv}
\widehat{RV}_{t,h}^{(JWTSRV)}=\sum_{j=1}^{J^m+1}\widehat{RV}_{j,t,h}^{(JWTSRV)}=\sum_{j=1}^{J^m+1}\left(\widehat{RV}_{j,t,h}^{(W,J)}-\frac{\bar{N}}{N} \widehat{RV}_{j,t,h}^{(WRV,J)}\right),
\end{equation}
where $\widehat{RV}_{j,t,h}^{(W,J)}=\frac{1}{G} \sum_{g=1}^G  \sum_{k=1}^N \mathcal{W}_{j,t-h+{\frac{k}{N}h}}^2 $ is obtained from wavelet coefficient estimates on a grid of size $\bar{N}=N/G$ and $\widehat{RV}_{j,t,h}^{(WRV,J)}=\sum_{k=1}^N \mathcal{W}_{j,t-h+{\frac{k}{N}h}}^2$ is the wavelet realized variance estimator at a scale $j$ on  the jump-adjusted observed data, $y_{t,h}^{(J)}$.\\

The JWTSRV estimator decomposes the realized variance into an arbitrary chosen number of investment horizons and jumps. \cite{barunik} discuss that it is consistent estimator of the integrated variance as it converges in probability to the integrated variance of the process $p_t$. \cite{barunik} also test the small sample performance of the estimator in a large Monte Carlo study and they find that it is able to recover true integrated variance from the noisy process with jumps very precisely. They also run a forecasting simulation where JWTSRV estimator confirms to improve forecasting of the integrated variance substantially.

\subsection{Model-free implied volatility and corridor implied volatility}

While realized volatility measures the volatility from the high frequency returns, model-free implied volatility (MFIV) can be used to infer the volatility from the option prices. This approach, derived by \cite{BN00} uses cross-section of option prices to calculate the volatility as the risk-neutral expected sum of squared returns between two dates. The resulting implied volatility does not depend on any parametric model and provide ex-ante risk-neutral expectations of the future volatilities. The most serious forerunner of MFIV was the volatility inverted from the \cite{BS73} option pricing formula. Nevertheless it has been proved that Black-Scholes implied volatility featured a notarially known moneyness  bias, known as volatility smile or smirk (\cite{MM79} were amongst the first researches to describe this issue). 
 
\cite{BN00} derived the model-free implied volatility under the diffusion assumption.  They extended the work of \cite{DK94} and \cite{D94} to infer a forecast of underlying asset's volatility from a continuum of European call options with strikes and maturities ranging from zero to infinity. The complete set of option prices is used to extract a condition which characterizes the set of continuous processes consistent with current option prices.

In the option setting, current time is fixed to $t=0$, pay-off takes place at a future fixed date $T$, time to maturity is denoted as  $\tau=T-t$. For $0 \leq t \leq T$, $F_t$ denotes the time $t$ value of the futures contract expiring at date $T\sp{\prime} $. Prices of European put and call options with strike $K$ and expiration date $T$ are given by $P_{t}(K)$ and $C_{t}(K)$ and risk-free rate is assumed to be zero.

Authors first derive the risk-neutral probability of the stock price that is fully determined by initial set of option prices. Consequently they show that the set of initial option prices determine as well the probability of the stock price reaching any two price levels at two consecutive dates. \cite{Jiang2005} further extend \cite{BN00} volatility measure to all martingale asset price processes and express the formulae in forward rather than spot prices. Since a martingale can be decomposed canonically into the orthogonal sum of a purely continuous martingale and a purely discontinuous martingale \citep{JS87,P90}, the model-free relationship between the asset return variance and option prices also holds when asset prices contain jumps.

A forecast of integrated variance for the period $[0,T]$ can be determined from observed European call option prices with maturity $T$ as follows
 \begin{equation}
 \label{MFIV}
\ E_0^F \left[\int_0^T \sigma^2_s ds \right ] = 2 \int_0^\infty \frac{C_t\left (T,K \right) - \max\left ( 0,F_{0}-K \right ) }{K^{2}} dK
\end{equation}
where $E_t^F$ denotes the time $t$ expectation with respect to the risk-neutral distribution (RND) of the asset price, $T$ denotes expiration date, $F$ forward probability measure, $K$  strike price and $F_{t}$ and $C_t \left (T,K \right )$ are forward asset price and forward option price respectively. To calculate the integral, numerical integration can be used, e.g. trapezoidal rule
\begin{equation}
\int_{K_{min}}^{K_{max}} \frac{C_t\left (T,K \right ) - \max\left ( 0,F_{0}-K \right ) }{K^{2}} dK\approx \sum_{i=1}^{m}\left [g\left (T,K_i\right)+g\left (T,K_{i-1}\right)\right ]\Delta K,
\end{equation}
where $\Delta K=\left(K_{max}-K_{min}\right)/m$, $K_i=K_{min}+i\Delta K$, and $g\left(T,K_i\right)= [C_t\left (T,K \right ) -$ $\max \left(0,F_{0}-K \right)]/{K^{2}}$. 

\cite{Jiang2005} further developed, based on examination of implementation issues, a simple method to implement the MFIV on observed option prices. They identify two types of errors associated with implementation: truncation and discretization errors. Truncation errors are present when tails in the RND are ignored (due to limited availability of the strike prices for listed options). Authors find that truncation errors are negligible if RND is truncated at two standard deviations from $F_0$ and propose the flat extrapolation scheme for the range of strike prices outside the available set of the prices. In their later work (\cite{JT07}) authors propose to impose smooth pasting condition at the minimum and maximum of available strike prices to avoid kinks in the implied volatility function at the lower and upper price bound. We discuss both schemes later on in the data section. Discretization errors are minimized using interpolation between listed strike prices. Cubic spline interpolation is not applied directly to option prices as there is nonlinear relationship between the option prices and option strike prices. Implied volatilities are obtained from \cite{BS73} formula, smooth function is fitted to implied volatilities and using \cite{BS73} formula volatilities are again translated to option prices at the desired strike prices. The \cite{BS73} model works as one-to-one mapping between volatilities and option prices and does not impose any model dependency on the calculation of model-free volatility.

\cite{Andersen2007} and \cite{Andersen2011} argue that lack of the liquid contracts results into a non-trivial measurement errors that are amplified by the stochastic nature of availability of strike prices which vary over time.  Authors propose corridor implied volatility measure following the work of \cite{CM98} with cut-off criterion that is determined endogenously by option prices (part of the estimated risk neutral density inferred from option prices). As such, it allows to reflect the pricing of volatility across an economically equivalent fraction of the strike range. That would ensure inter-temporal coherence of the measure. Authors further show that due to lack of availability of strike prices, implementation of MFIV actually brings the resulting implied volatilities to corridor-implied volatilities rather than model-free implied volatilities.

By defining two positive barriers, the lower $B_1$, the upper $B_2$ and the following indicator function, 
\begin{equation}
I_t(B_1,B_2)=I_t=1\left[B_1\leq F_t\leq B_2\right],
\end{equation}
corridor integrated variance is 
\begin{equation}
CIVAR_t (B_1,B_2)=\int_0^T\sigma_s^2I_s(B_1,B_2)ds.
\end{equation}
In comparison to the pure integrated variance, now the return variation is accumulated only when the futures price at time $t$ is between the two barrier levels. \cite{CM98} and \cite{Andersen2007} demonstrate that risk-neutral expectation of future corridor implied variance (CIV )at time $t=0$ is
\begin{equation}
\ E_0^F \left[\int_0^T \sigma^2_s I_s(B_1,B_2) ds \right ] = 2 \int_{B_1}^{B_2} \frac{C_t\left (T,K \right) - \max\left ( 0,F_{0}-K \right ) }{K^{2}} dK
\end{equation}
When $B_1=0$ and $B_2=\infty$ the future corridor integrated variance is equivalent to model-free implied variance. Expected future volatility can be obtained by taking square roots of CIV as well as MFIV.


\section{Data}

To calculate option implied volatility we use the set of European-style option prices on DAX and S\&P 500. The cross-sectional data contain daily option prices for all listed maturities and strike prices and cover the period from July 2006 till October 2010. The data have been provided by by OptionMetrics database. We use settlement mid price for the analysis. Settlement prices have the advantage over close prices as they are not plagued by nonsynchroneous trading. Following  \cite{BCH97}, we apply exclusions filters on the datasets to prevent liquidity related bias and mitigate the impact of price discreteness. First, options with less than week to expiration are excluded. Second, price quotes lower than 0.375 are excluded. Third, the quotes that do not satisfy the no-arbitrage condition $C\ge \max{(0,S_{t}-X_{t})}$ for calls and $P\ge \max{(0,X_{t}-S_{t})}$ for puts are dropped as well. We include only out-of-the money options where strike price is strictly higher then the spot price for call options and vice versa for put options. This approach is common in all comparable studies as in-the-money options are less liquid and thus introduce bias into the calculation of implied volatility.  Last, we calculate the measures if and only if more then five options for given maturity and different strikes passed the above mentioned criterions. Altogether, we discarded 27\% of option prices and used in average 80 options per day per maturity per index.

To deal with the fact that measures work with spot price and options are written on index forwards, we follow the literature and translate spot prices into forward rates using zero coupon rates for a given currencies and maturities for each day. In case we do not have zero coupons for given maturity, we interpolate (extrapolate) between the two nearest rates available. Zero coupon rates are again obtained from OptionMetrics database. To obtain forward prices, spot prices are multiplied by $e^{r_f(T-t)}$, where $r_f$ is the risk-free rate corresponding the maturity of options.

To calculate the model-free measures we used the datasets with fixed time to maturity; precisely, we use 15 and 28 calendar days for DAX, 15 and 29 for SPX. Note that the one-day difference is caused by different listing conditions for the index options. We refer to the different time to maturity as monthly and bi-weekly maturity further in the text. We use non-overlapping datasets using the data from the trading days when options with fixed time to maturity have been available so we do not introduce any autocorrelation into the forecasting regressions. When dealing with implementation of model-free implied volatility on real option datasets, one has to inevitably resort to an  approximation method introducing potential bias. \cite{Jiang2005} point out to some practical limitations associated with the implementation of MFIV which result to truncation errors when the tails of distribution is ignored, and discretization errors due to numerical integration and limited availability of strike prices range. Authors as well propose the procedure that improve the implementation and render more or less negligible errors.

In our calculations, we follow the part of this procedure. To limit the discretization errors, we use step of one unit of the index to calculate the integral numerically. Trapezoidal rule is applied to obtain the variance (correspondingly square root to obtain volatility). To overcome the limited availability of strike prices, we infer the prices of absent options from the prices of listed options. Due to nonlinearities in the option prices we apply cubic spline curve-fitting method to implied volatilities, instead of option prices directly. The implied volatilities are reverted from listed option prices using the  \cite{BS73} pricing formula and smooth function is fitted to them. Implied volatilities for absent strikes are then extracted and the same pricing formula is used again to calculate the corresponding option prices. The pricing formula is used only as one-to-one mapping between option prices and strike prices. Thus the procedure retains its model-free grounds. 

The subsequent procedure differs for MFIV and CIV calculation. We impose flat extrapolation to the prices beyond the available daily price range in MFIV measure implementation.  \cite{JT07} adjust the slope of the extrapolated segment  to match the corresponding slope of the interior segment at the extremes. We did not apply this approach as it actually rendered much worse results. The number of listed strikes that passed all above mentioned criterions varies between the days. In some cases the slope at the interior segment would send the extrapolated prices to unrealistic numbers, thus introducing huge bias into the calculated volatility. To avoid truncation errors we use one standard deviation from forward prices as an integration range. We do not follow \cite{Jiang2005} recommended approach to set the truncation point at two standard deviations from $F_0$ for simple reasons: the lack of available options for some days and high volatility of index prices through the time period that involves financial crisis. Interval of strike prices that then needs to be extrapolated $\left[ F_0-2SD,K_{\min}\right] \cup \left[ K_{\max},F_0+2SD \right]$) becomes too large and amounts for the majority of the inputs that enters the calculation formula, which necessarily introduces great amount of error into the calculation. We calculate CIV with the corridors covering range from the  5th to 95th percentile (CIV1) and from 2.5th to 97.5th percentile (CIV2) of the RND, estimated from the available option prices that passed the exclusion criteria for given daily maturity. 

Corresponding to each of the implied volatilities we compute the realized volatility (RV) using a 5 minute returns and JWTSRV using all the available data. Tick by tick data were provided by the Tick Data. After computing the daily realized measures, we aggregate them into monthly and bi-weekly according to the maturities in order not to introduce the bias from the maturity mismatch.



\section{Time-frequency volatility dynamics}

The information content of implied volatility is typically assessed in the literature by estimating a following regression
\begin{equation}
\label{IVRVOLS}
RV_{t+h}=\alpha+\beta IV_t+\epsilon_t,
\end{equation}
with ordinary least squares assuming $\epsilon_t$ to follow Gaussian normal errors with zero mean and finite constant variance. $RV_{t+h}$ is the ex-post realized volatility for period $t+h$ and $IV_t$ is the implied volatility at the beginning of period $t$, being an ex-ante measure of $t+h$ volatility. In case implied volatility is an unbiased forecast of realized volatility, $\alpha$ should not be statistically different from 0 and $\beta$ should not be statistically different from 1. In case implied volatility is efficient, the residuals $\epsilon_t$ should be zero mean, constant and finite variance and they should be serially uncorrelated.

Initial literature have generally found that implied volatility is biased forecast of the realized volatility while $\beta$ is significantly different from unity, see for example \cite{christensen1998}.  A few researchers \citep{Bandi2006,Christensen2006} suggest that the implied-realized volatility relation might be a fractional cointegration relation as volatility is typically found to be a long memory process. In this case, $\epsilon_t$ would not be integrated of order $I(0)$ and standard OLS should not be used. Before proceeding further with the cointegrating relationship, we study the time-frequency dynamics of the relation using wavelet coherence to find out how the dependence varies over different frequencies. This will provide an important insight for the further analysis.

\subsection{Dynamic dependence: a wavelet coherence}

To better understand the relationship, it might be useful to look at it from the point of view of different frequencies. Here, the wavelet analysis may be well utilized as it allows to study the time series in the time-frequency domain. As wavelet coefficients estimate the spectrum of the time series, wavelet coherence can be seen as the estimate of the cross-spectra between two series scaled by the spectra of both series. The coherence is analogous to the square of the correlation between two series. Zero coherence suggests that there is no relation, when coherence equals to one, we have perfect correlation. The main advantage of this approach is that it provides us with the localized correlation at time-frequency domain. As such, it can be used as a ``lens" into the dependence. 

The wavelet transform offers localized frequency decomposition, providing information about frequency components. As a result, wavelets have significant advantages over basic Fourier analysis when the object under study is locally stationary and inhomogeneous -- see \cite{Gencay2002, PercivalWalden2000,Ramsay2002}.

To be able to study the interaction between two time series, we use a bivariate framework of wavelet coherence\footnote{In our work we use continuous wavelet analysis tools, for an interested reader, we include the necessary introduction to wavelet analysis in an Appendix \ref{appWavelets}}. Following \cite{TorenceCompo98}, we define the cross wavelet transform of two time series $RV_{t+h}$ and $IV_t$ as
\begin{equation}
W_{IV_tRV_{t+h}} (u,s) = W_{IV_t} (u,s) W_{RV_{t+h}}^* (u,s),
\end{equation}
where $W_{IV_t} (u,s)$ and $W_{RV_{t+h}} (u,s)$ are continuous wavelet transforms of $RV_{t+h}$ and $IV_t$, respectively, $u$ is a position index, and $s$ denotes the scale, while the symbol $^*$ denotes a complex conjugate. The cross wavelet power can easily be computed using the cross wavelet transform as $|W_{IV_tRV_{t+h}} (u,s)|$. The cross wavelet power reveals areas in the time-frequency space where the time series show a high common power, i.e., it represents the local covariance between the time series at each scale.

The wavelet coherence can detect regions  in the time-frequency space where the examined time series  co-move, but do not necessarily have a high common power. Following the approach of \cite{TorrenceWebster99}, we define the squared wavelet coherence coefficient as: 
\begin{equation}
R^2 (u,s)=\frac{|S(s^{-1}W_{IV_tRV_{t+h}} (u,s))|^2}{S(s^{-1}|W_{IV_t} (u,s)|^2) S(s^{-1}|W_{RV_{t+h}} (u,s)|^2)},
\end{equation}
where $S$ is a smoothing operator\footnote{Without smoothing, the wavelet coherence equals one at all scales. Smoothing is achieved by convolution in both time and scale. The time convolution is performed with a Gaussian window, while the scale convolution is done with a rectangular window -- see \cite{Grinsted2004}.}. The squared wavelet coherence coefficient is in the range $0\le R^2 (u,s) \le1$. Values close to zero indicate weak correlation, while values close to one provide evidence of strong correlation. Hence, the squared wavelet coherence measures the local linear correlation between two stationary time series at each scale and is analogous to the squared correlation coefficient in linear regression. Since the theoretical distribution for the wavelet coherence is not known, statistical significance of dependence is tested using Monte Carlo methods \citep{Grinsted2004,TorenceCompo98}\footnote{The use of wavelets brings with it the difficulty of dealing with boundary conditions on a dataset with finite length. This is a common problem with any transformation relying on filters. In our paper, we deal with this problem by padding the time series with a sufficient number of zeroes. The area where the errors caused by discontinuities in the wavelet transform cannot be ignored, i.e., where edge effects become important, is called the cone of influence. The cone of influence is highly dependent on the type of wavelet used -- see \cite{TorenceCompo98}. The cone of influence lies under a cone which is bordered by a thin black line.}.

Finally, wavelet coherence phase differences may be used to asses the details about the delays in the oscillation (cycles) between the two time series under study (see \cite{TorrenceWebster99} for the details). Phase is indicated by arrows on the wavelet coherence plots. A zero phase difference means that the examined time series move together. The arrows point to the right (left) when the time series are in-phase (anti-phase) or are positively (negatively) correlated. Arrows pointing up means that the first time series leads the second one by $90^{\circ}$, whereas arrows pointing down indicates that the second time series leads the first one by $90^{\circ}$. Usually we have a mixture of positions, for example, an arrow pointing up and right means that the time series are in phase, with the first times series leading the second one. Recently, \cite{vachabarunik} used the wavelet coherence to study the comovements of commodity markets. 

\begin{figure}
   \centering
   \includegraphics[width=3.2in]{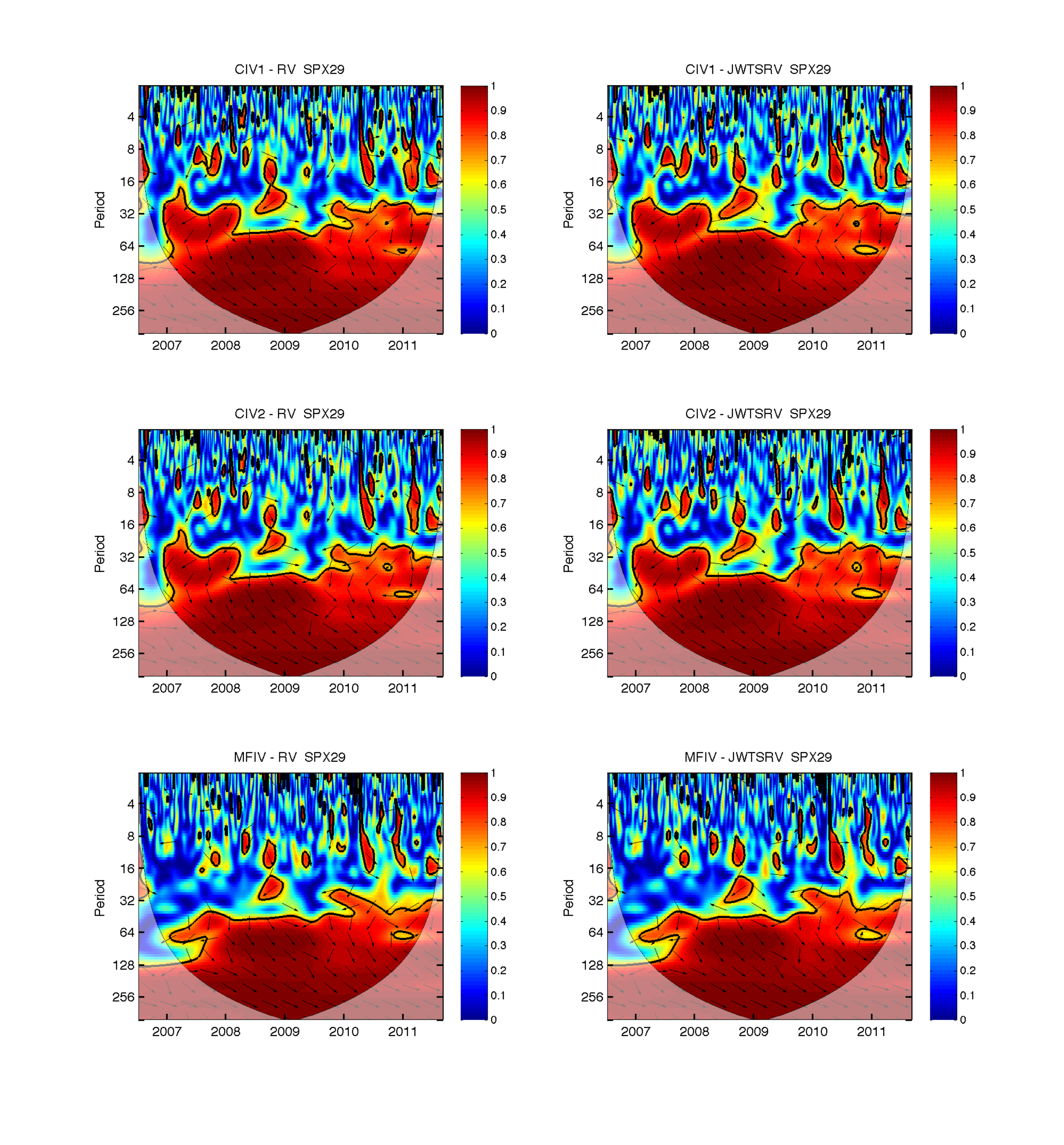} 
   \includegraphics[width=3.2in]{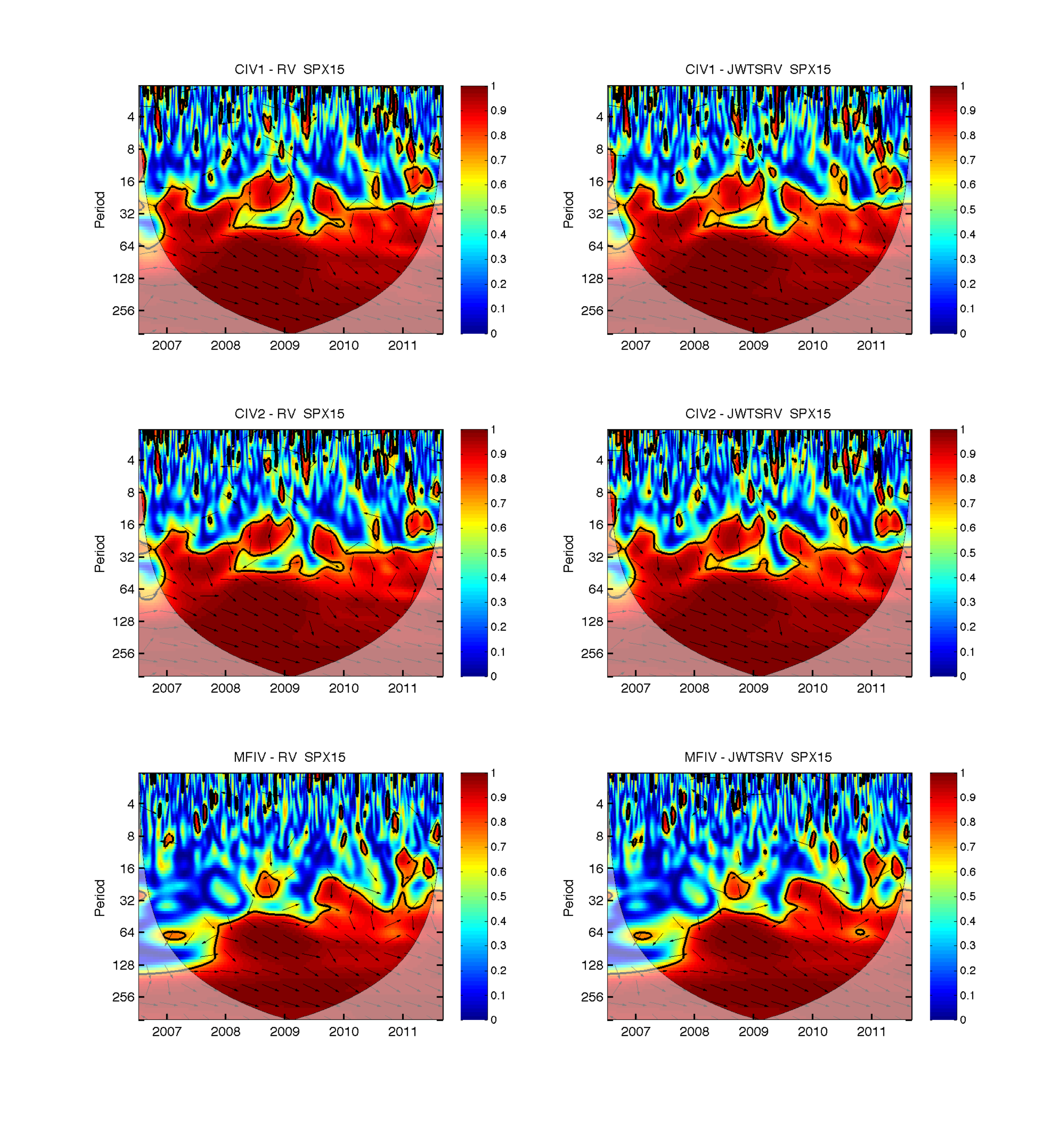} 
      \includegraphics[width=3.2in]{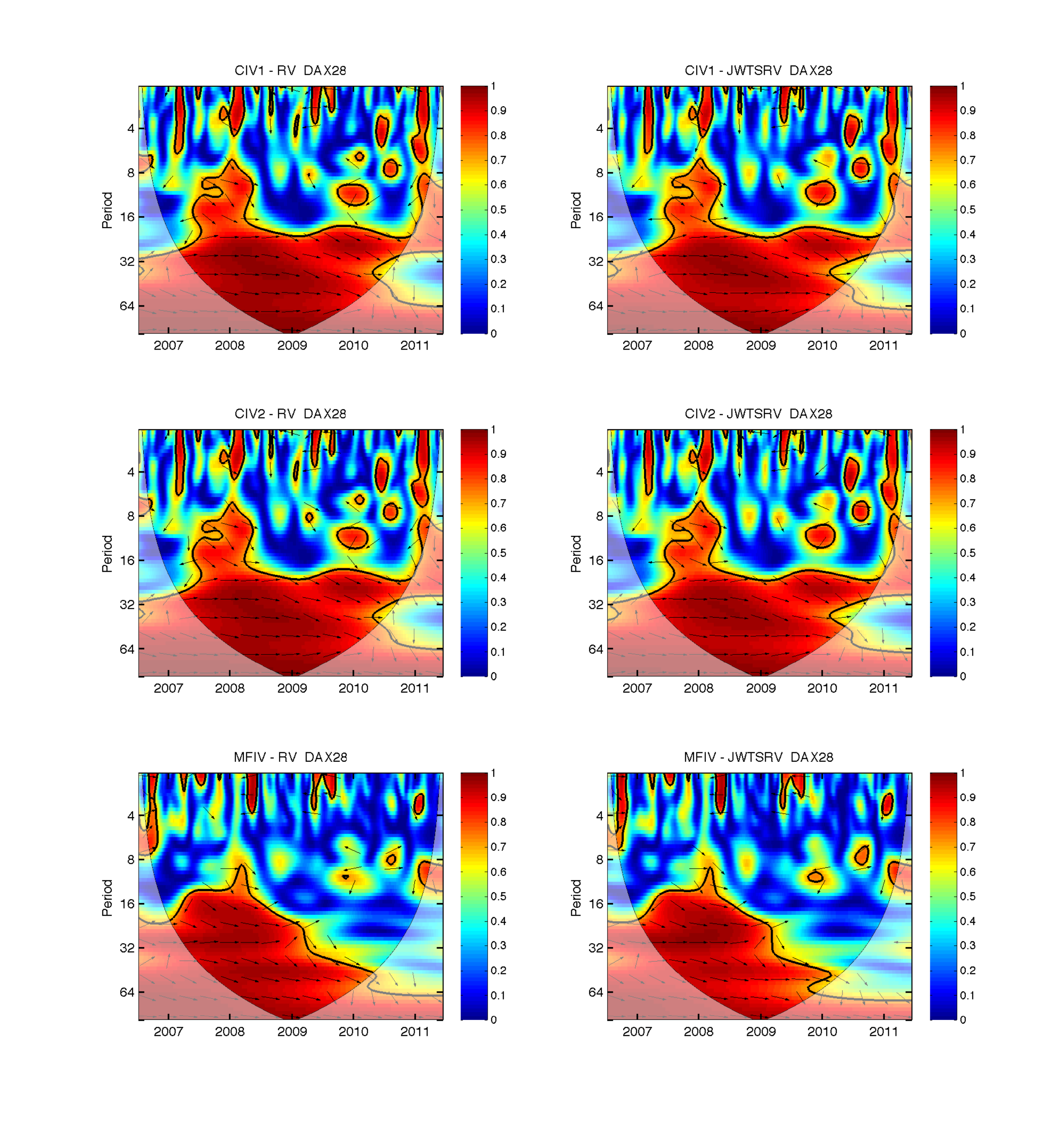} 
    \includegraphics[width=3.2in]{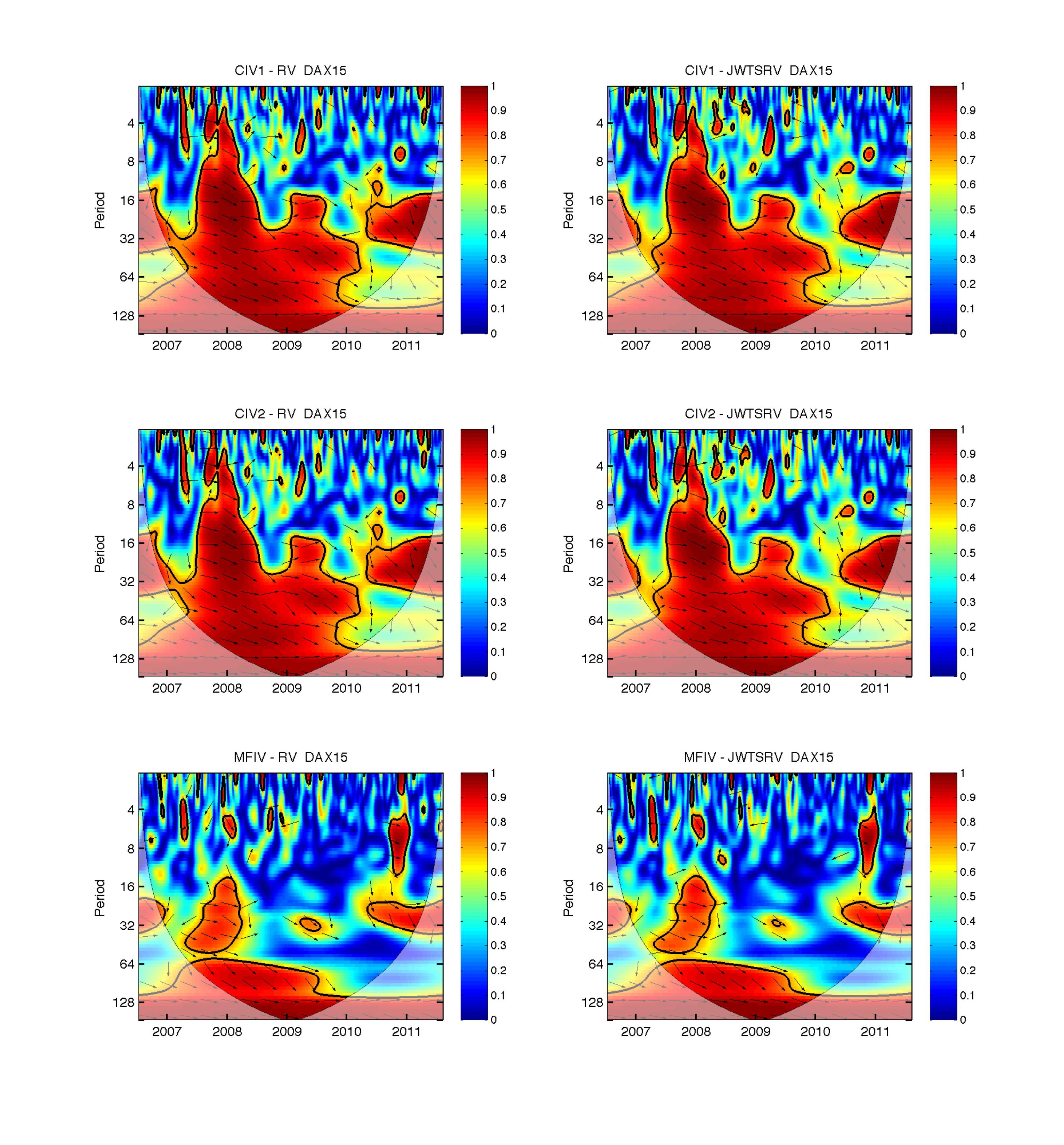} 
   \caption{S\&P 500 and DAX monthly and bi-weekly wavelet coherences between implied and realized volatility measured by CIV1, CIV2, MFIV and RV and JWTSRV respectively.}
   \label{wtcSPX}
\end{figure}

Figure \ref{wtcSPX} brings the wavelet coherence plots of the two variance series. We use CIV1, CIV2 as well as MFIV to measure volatility implied by options and RV and JWTSRV robust to noise and jumps to measure the realized volatility. Moreover, for each index, we use options with monthly and bi-weekly maturity to study the difference. Realized volatility is computed correspondingly.

Dynamic dependence reveals an interesting findings. A distinct change in the general pattern can be found in the $2^5$ frequency corresponding to 32 days, or approximately 1.5 months when 21 trading day is considered in a month. In the investment horizons less than 1.5 months, no dependence is found, while the wavelet coherence is significant through horizons longer than 1.5 month and all time periods considered. Thus effects of market frictions and short-run fluctuations disappear in the long-run and dynamic relationship between the variances is nearly perfect in the long-run for the whole studied period. In the long-run (low frequencies) coherence close to one implies that implied volatility is an unbiased forecast of realized volatility and no forecast error or premium for bearing volatility risk exist. In the short run this equilibrium is broken and zero coherence implies that most of the changes in implied volatility is coming from the risk premium or errors in future expectations.

Interestingly, options with different maturities used to calculate the implied volatility of S\&P 500 do not bring any difference into the relation. Implied-realized volatility relation is very strong for all investment horizons longer than 1.5 months and time periods without an exception. The situation is similar in the DAX. The only difference is that relation seems to break in the last years of the sample even in the long horizons. This may suggest that German option market is not that efficient as the U.S. one.  

An important distinction can be seen from the wavelet coherence plots when we consider different measures of volatility used. When CIV measures are used to calculate the implied volatility, the long-term relation is much stronger than in the case when MFIV is used. This may suggest that MFIV provide biased measure of implied volatility as long run dynamics of the relation is not so pronounced. 

Last observation can be made when looking at phases (arrows in the plot) which point down-right. This means that implied volatility generally leads the realized volatility, which is expected in case implied volatility provides an efficient expectation about future volatility.

While wavelet coherence plots provide us the ``lenses" into the implied-realized relation, in the next sections we will develop rigorous methodology to estimate the long-term fractional cointegration relation using wavelets. Our main motivation in doing so is that wavelets are capable of dealing with non-stationary time series, which will turn to be crucial property for the analysis.

\subsection{Fractional cointegration in variances}

Fractional integration provides a framework for studying long-run dependencies in economic time series \citep{bailie1996}. A stationary time series $y_t$ is said to be fractionally integrated of order $d\in(0,0.5)$, $I(d)$, if 
\begin{equation}
\Delta^dy_t=\epsilon_t,
\end{equation}
where $\epsilon_t$ is an $I(0)$ process and $\Delta^d=(1-L)^d$ is fractional difference operator. Empirical evidence suggests that financial market volatilities are well described by the $I(d)$ processes \cite{abde2001,abdl2003,Christensen2006}. Naturally, we can expect that the implied and realized variance series will be tied together in the long-run relationship in the form of fractional cointegration and a linear combination of the two will be integrated of order lower than $I(d)$. Thus the difference between implied and realized variance, variance risk premium, should be less persistent that the two individual variance series. This result has been documented by \cite{Christensen2006}. Interestingly, \cite{Bandi2006,Kellard2010} report fractional order of volatility in a non-stationary region when $1/2<d<1$, although it is difficult to determine the integration order of fractional variables as smooth transition exists between stationary and non-stationary regions \citep{marinuccirobinson2001}.

When looking at the regression Eq. (\ref{IVRVOLS}), for $\alpha=0$ and $\beta=1$, residuals $\epsilon_t$ reduces to a variance risk premium obviously. In case $\epsilon_t$ is an $I(d_u)$ with $d_u<d$, we may suspect the fractional cointegrating relation between implied and realized variances. 

\subsection{Band spectral regression approach}

Part of the literature proposes to use a band spectral regression to estimate a fractionally cointegrating relation in implied and realized volatilities, as OLS estimates of $\beta$ are inconsistent. \cite{robinsonmarinucci2003, Christensen2006} have shown that narrow band least squares (NBLS) results in an estimator that is consistent and normally distributed. Basic idea is transforming the time series into the frequency domain using Fourier transform and estimating $\beta$ on the narrow band of the spectrum (Fourier coefficients) not far from the zero frequency on the long memory region. Recently, \cite{NielsenFrederiksen} generalize this idea to a fully modified NBLS (FMNBLS) which is able to deal with the bias introduced by correlation between regressors and errors. In this paper, we build on these ideas, but use rather different approach of band least squares on the spectra estimated on wavelet coefficients.

Let us introduce the approach by considering the regression model
\begin{equation}
y_t=x_t\beta+\epsilon_t,
\end{equation}
where $\{x_t,t=1,\ldots,T\}$, $\{y_t,t=1,\ldots,T\}$ and $\epsilon_t\sim N(0,\sigma^2)$. The OLS estimator of $\beta$ is
\begin{equation}
\hat{\beta}^{OLS}=\left(x^Tx\right)^{-1}x^Ty=\frac{Cov(x_t,y_t)}{Var(x_t)}.
\end{equation}
\cite{engle1974} was among first to consider estimation of $\beta$ in the frequency domain. In fact frequency domain is very intuitive as variance and covariance are spectrum and co-spectrum of the series and can be simply estimated for example using Fourier transform.

Recently, \cite{Kellard2010} finds that realized as well implied volatility series may lie in the non-stationary region when $1/2<d<1$. Frequency domain least squares using the Fourier transform are able to accommodate non-stationary fractional cointegration by transforming potentially non-stationary series $x_t$ which are $I(d)$ with $d>1/2$ using $\gamma \ge0$ into the resulting $\Delta^{\gamma}x_t$ which are $I(d-\gamma)$ \citep{NielsenFrederiksen}. The choice of $\gamma$ affects the estimation procedure and different choices will lead to different estimators. Authors propose the best choice of $\gamma=d_u$, where $d_u$ is memory parameter of the residuals which can be estimated. 

Frequency domain least squares based on the wavelet estimation of spectra is able to deal with this problem as wavelets are generally very convenient tool in case we are dealing with the non-stationary series \citep{fanwhitcher,Roueff2011}. Although wavelets do not improve the estimation of $d$ in the standard stationary context $d < 1/2$, \cite{faymoulines2009} showed that in the presence of trends, or series with $d \ge 1/2$ and $d \le -1/2$, they are helpful as they allow differencing implicitly.

\subsection{Wavelet band spectral regression (WBLS)}

Using wavelet transform, we are able to divide the whole frequency spectrum into frequency bands represented by wavelet scales $j$. After the transform, the resulting spectrum on the $j$-th scale has the following form: $f^W_j\in[1/2^{j+1},1/2^{j}]$. The wavelet spectral density function at a scale $j$ can be expressed as $S_{(x)j}(f)=\mathcal{H}_j(f) S_x(f)$ where $\mathcal{H}_j(f)$ is the transfer function of the wavelet filter at a scale $j$ and $S_x(f)$ denotes the spectrum of $x_t$. Similarly the wavelet cross-spectrum at a scace $j$ is defined as $S_{(xy)j}(f)=\mathcal{H}_j(f) S_{xy}(f)$, where $S_{(xy)j}(f)$ represents the cross-spectrum of $x_t$ and $y_t$. Furthermore, wavelet variance $\nu^2_x(j)$ and wavelet covariance $\gamma_{xy}(j)$ at a scale $j$ reads:
\begin{equation}
\nu^2_x(j)=\int_{-1/2}^{1/2} S_{(x)j}(f) df = \int_{-1/2}^{1/2} \mathcal{H}_j(f) S_x(f) df,
\end{equation}
\begin{equation}
\gamma_{xy}(j)=\int_{-1/2}^{1/2} S_{(xy)j}(f) df = \int_{-1/2}^{1/2} \mathcal{H}_j(f) S_{xy}(f) df,
\end{equation}
In case $T\rightarrow \infty$ and therefore maximum number of wavelet scales $J\rightarrow \infty$ is available, we can write total  variance and covariance as a sum of wavelet variances and covariances at all scales as \citep{WGPTech1999}:  
\begin{equation}
Var(x_t)=\left(x^Tx\right)=\int_{-1/2}^{1/2} S_{(x)j}(f) df = \sum_{j=1}^{\infty}\nu^2_x(j)
\end{equation}
\begin{equation}
Cov(x_t, y_t)=\left(x^Ty\right)=\int_{-1/2}^{1/2} S_{(xy)j}(f) df = \sum_{j=1}^{\infty} \gamma_{xy}(j)
\end{equation}

Thus $\beta$ can be estimated in the frequency domain using wavelet least square estimator (WLS) as follows
\begin{equation}
\label{WLS}
\hat{\beta}^{WLS}(1,\infty)=\left(\sum_{j=1}^{\infty} \gamma_{xy}(j) \right) \left(\sum_{j=1}^{\infty}\nu^2_x(j)\right)^{-1}
\end{equation}
Asymptotically, $\hat{\beta}^{WLS}$ is equal to $\hat{\beta}^{OLS}$. In many situations time series cary different information in the low and high part of the spectra. As revealed by the wavelet coherence, this is also the case for volatility relation. As the relation comes solely from the long-run part of the spectra, we need a tool which will be able to estimate the relation only on this part. Similarly to the NBLS and FMNBLS \citep{robinsonmarinucci2003, Christensen2006,NielsenFrederiksen}, which we introduce later in the text as we use it for comparison with our estimator, we can obtain the estimate on the narrow band of the spectrum not far from the zero frequency on the long memory region. More precisely, we can use only scales $j$ which cover the long memory region.

Our final estimator, wavelet band least square estimator (WBLS) simply estimates the $\beta$ on the band of scales $j\in[k,l]$ in Eq. (\ref{WLS}), thus using frequency band $f\in[1/2^{l+1},1/2^{k}]$. For the estimation of spectra, we use modified discrete wavelet transform (MODWT)\footnote{For a more detailed treatment see appendix}. WBLS estimator is then 
\begin{equation}
\label{betaw}
\hat{\beta}^{WBLS}(k,l)=\left(\sum_{j=k}^{l} \gamma_{xy}(j) \right) \left(\sum_{j=k}^{l}\nu^2_x(j)\right)^{-1} 
\end{equation}
where $j$-th scale represents the frequency band $f\in[1/2^{j+1},1/2^{j}]$. For example $\hat{\beta}^{WBLS}(3,4)$ will estimate $\beta$ over the frequency band of $f\in[1/2^5,1/2^3]$. The estimator in Eq. (\ref{betaw}) can be expressed in terms of the MODWT coeffiecients $\widetilde{W}_{x}(j,s)$ and $\widetilde{W}_{y}(j,s)$, where $j$ and $s$ denote scale position of the transform for $x_t$ and $y_t$ as

\begin{equation}
\hat{\beta}^{WBLS}(k,l)=\left(\sum_{j=k}^{l} \left[\frac{1}{T}\sum_{u=1}^T\widetilde{W}_{x}(j,s) \widetilde{W}_{y}(j,s)\right]\right) \left(\sum_{j=k}^{l} \left[\frac{1}{T}\sum_{u=1}^T\widetilde{W}^2_{x}(j,s)\right]\right)^{-1} 
\end{equation}


By WBLS we can focus on estimating long-memory part of the spectra using frequency band near origin and obtain long memory relationship. When we use all scales $j$ and as $j\rightarrow\infty$, $\hat{\beta}^{WBLS}$ will be equivalent to $\hat{\beta}^{WLS}$ and will converge to an OLS estimator. 

\subsection{FMNBLS}
For the comparison with the newly proposed WBLS estimator of fractional cointegrating implied-realized relation, we use frequency domain least squares methods well-established in the literature. The basic distinction from the WBLS is that instead of using wavelet coefficients to estimate the spectra and co-spectra, Fourier transform is used by the rest of the literature.

Basic idea of the narrow band least squares (NBLS) estimator  is transforming the time series into the frequency domain using Fourier transform and estimating $\beta$ on the narrow band of the spectrum not far from the zero frequency on the long memory region. \cite{robinsonmarinucci2003, Christensen2006} have shown that narrow band least squares (NBLS) results in an estimator that is consistent and normally distributed.  Averaged (co-) cross-periodogram used for the estimation of spectrum is $\hat{F}_{xy}(k,l)=2\pi/T\sum_{j=k}^{l}I_{xy}(\lambda_j)$ for any $0\le k \le l \le T-1$ and for $\lambda_j=2\pi j/T$, where $I_{xy}(\lambda_j)=1/2\pi T\sum_{t=1}^{T} \sum_{s}^{T}x_t y_t'e^{-i(t-s)\lambda}$ is cross-periodogram, or estimated cross-spectrum between two series on a specific frequency band $[k,l]$.  Analogously, $\hat{F}_{x}(k, l)$ is estimated spectrum of $x_t$. Then the cointegrating relation between two time series $\{x_t\}$ and $\{y_t\}$ can be estimated as
\begin{equation}
\hat{\beta}^{NBLS}(k,l)=\hat{F}_{xy}(k,l)\hat{F}_{x}^{-1}(k,l),
\end{equation}
where $k$ and $l$ define the frequency band used for the estimation of $\beta$.

By definition, $\hat{\beta}^{NBLS}(1,T-1)$ is algebraically identical to usual OLS estimator of $\beta$ and thus identical to WLS estimator in Eq. (\ref{WLS}). If $\frac{1}{l}+\frac{l}{T}\rightarrow0$ as $T\rightarrow\infty$, $\hat{\beta}^{NBLS}(k,l)$ is an NBLS estimator using only degenerating band of frequencies near the origin. While $l$ needs to tend to infinity to have information, it also needs to remain in a neighborhood of zero where we have assumed knowledge about the spectral density, so $l/T$ must tend to zero.

\cite{NielsenFrederiksen} based on their previous work \citep{Christensen2006} show that in absence of non-coherence between regressors and errors at zero frequency imposes bias to the NBLS estimates and they propose a fully modified NBLS (FMNBLS) estimator to eliminate  this bias. The FMNBLS estimator is simply NBLS corrected for the asymptotic bias estimated by running an auxiliary NBLS regression of the (differenced) residuals from the initial NBLS on the same regressors. To keep the text under control, we point interested reader to the work of \cite{NielsenFrederiksen} for details of the methodology.


\section{Results}

The main aim of the paper is to revisit the relationship between implied and realized volatility using the new unbiased measures of volatilities and newly proposed wavelet band spectral regression. 

Wavelet coherence suggests that implied volatility may be an efficient forecast of the future volatility in the long run, while the existence of risk premia makes it an inefficient forecast in the short run.  Until now, we have been using the daily data as wavelets are known for their decorrelation properties and they can deal with the non-stationary data as well. \cite{christensen1998} were first to note that overlapping data may affect the estimation considerably. To overcome this problem, we aggregate the daily data to a monthly (and bi-weekly) non-overlapping data for further analysis. 

We begin to study the volatilities by estimating its long memory parameter. Table \ref{GPH_raw} shows the memory estimates of different implied and realized volatility measures used in our study. For the estimation, we use the popular GPH method \citep{gph1983,robinson1995} and we report the estimates up to various different frequencies $T^{0.6}$, $T^{0.7}$ and $T^{0.8}$. Interestingly, nearly all estimated volatilities show memory lying in between the stationary and non-stationary region of $d=0.5$. When implied volatility is measured using corridor implied volatility (CIV1 and CIV2), it displays larger memory than conventional MFIV measure constituting the popular VIX index. Both realized volatility measures show similar memory which is uniformly lower than the memory of implied volatility. These results suggest that choice of the different measures may have serious consequence for the estimation of relation, especially due to the implied volatility measure. When compared to the literature which uses either model-driven implied volatilities or MFIV, our estimated implied volatilities have larger memory possibly crossing boundary of stationarity. It is worth to note that the period used in our study covers the recent financial crisis which could possibly introduce this feature.

As a preliminary step in the estimation of the relation, we estimate it using OLS. The results are reported together with frequency domain estimates separately for each series. Tables \ref{SPX29}, \ref{SPX15}, \ref{DAX28} and \ref{DAX15} report the OLS estimates for the S\&P 500 using monthly and bi-weekly maturities, DAX using monthly and bi-weekly maturities in the implied volatility measurement respectively. The first rows show the coefficients of the Eq. (\ref{IVRVOLS}) suggesting that generally IV is a biased predictor of RV. However, when CIV is used to measure the implied volatility, slope coefficients are strikingly closer to unity than MFIV estimates for both markets and both maturities used. This measurement error may have led researchers to find bias more pronounced than it may be. 

Memory estimates of the residuals from the OLS confirm this result. When MFIV is used, residuals are $I(d)$ suggesting implied-realized volatility is a cointegrating relation. When CIV is used as the implied volatility measure, except for the case of S\&P 500 with monthly maturity options, residuals does not display as pronounced long memory as reported in literature. In all the cases when residuals show long memory parameter close to zero, or $I(0)$, slope coefficient is very close to unity. 

While estimated volatilities tend to be at the boundary of stationarity, wavelet band least squares (WBLS) may serve as the best tool for the estimation of relation. From the wavelet coherence plots we can see that there is no relationship until the 32nd period, while the relationship for the periods higher than 32 seems to be nearly perfect. Thus we can utilize the result and conveniently estimate the relationship only on this part of the spectra using wavelet band least squares (WBLS).

Tables \ref{SPX29}, \ref{SPX15}, \ref{DAX28} and \ref{DAX15} report the results for the estimated $\beta$ using WBLS. We use $j=\{5,6\}$ levels from the 6 level wavelet decomposition to estimate the relationship. All the $\beta$'s are much closer to unity, regardless the measure used. It is interesting to note that when CIV is used, relationship is not significantly different from unity in most of the cases, while CIV1 measure implies coefficients closest to unity than does the CIV2. The only exception is the S\&P 500 data with monthly options where the coefficients are significantly lower than unity. MFIV provide the $\beta$ which is lower than unity all the times again. The difference between RV and JWTSRV measures is not so pronounced, although JSTWRV does bring some improvement in the estimates\footnote{As JWTSRV is relatively new measure, we have also used realized measures well established in the literature, namely bipower variation (BPV), realized kernels (RK) and the two-scale realized variance (TSRV) and the results are very similar to those reported here. To keep the page numbers under control, we do not report these results and make them available upon request from authors.}. This confirms the result of \cite{martin2011} who assess the robustness of the relative performance of various estimators to the microstructure noise and they find that results are invariant to the method of noise correction in the realized volatility.

For the comparison, we also estimate the FMNBLS. While in the WBLS we have motivated the choice of the bands by the wavelet coherence plots, in the FMNBLS we follow \cite{NielsenFrederiksen} and use the same bands to estimate the relationship, namely we use $[T^{0.4},T^{0.6}]$, $[T^{0.4},T^{0.7}]$, $[T^{0.4},T^{0.8}]$, $[T^{0.5},T^{0.6}]$, $[T^{0.5},T^{0.7}]$ and $[T^{0.5},T^{0.8}]$ frequencies. All of the bands overlap the bands used in the WBLS, although they interfere also with higher frequencies. The results are statistically similar to those obtained by the WBLS estimates. When CIV measures are used, $\beta$ is closer to unity (or significantly does not differ from unity) when compared to usual OLS. When MFIV is used, $\beta$ is closer to unity, but in most cases they are significantly lower than unity. 

Finally, it is worth to notice the difference between monthly and bi-weekly regressions. In the case of monthly regressions, when CIV1 measure is used, spectral band regressions confirm the long-run unbiasedness of the implied volatility forecasts. Residuals does not display significant long memory, thus fractional cointegration describes the relation well. On the other hand, unbiasedness of bi-weekly volatility forecasts is confirmed, but residuals from both WBLS as well as FMNBLS display significant long memory. OLS residuals does not show significant long memory and $\beta$s from OLS are very close to unity.
 
\section{Conclusion}
In this paper, we study the long-run unbiasedness of implied volatility as a predictor of future volatility. While this relation has been studied previously in the literature, our work contributes to the findings in several ways. First, we propose a new spectral techniques to estimate the potential fractional cointegrating relation of the implied and realized volatility based on wavelets. Main advantage in comparison to common spectral regression techniques based on the Fourier coefficients is that wavelets allow us to work with locally stationary series. Second, we study the fractional cointegration of the implied and relalized volatility using accurate corridor implied volatility (CIV measure) for the first time as most of the literature uses model based implied volatilities. 

When CIV is used to measure implied volatility on the options with monthly maturities, implied volatility is found to be an unbiased forecast of the realized volatility in the long term horizon over one month. This result holds for the options on S\&P 500 as well as DAX indices. Implied and realized volatility is confirmed to be fractionally cointegrating relation jointly using our newly proposed wavelet band least squares as well as fully modified narrow band least squares. In contrast, when MFIV is used as a measure of implied volatility, all estimates are lower than unity. This result strongly suggests that measurement of volatility implied by option prices is crucial for the volatility forecasts as wrong measurement introduce bias to the forecasts. The result is also important to the literature as it may suggest that estimated bias of option implied volatility forecasts may not be that pronounced as it is caused by the imprecise measurement.

We also question the importance of measures used on the other side of the regression testing the unbiasedness of implied volatility, namely realized volatility. Interestingly we find that results are invariant to the method of noise correction in the realized volatility.

\bibliography{CIV_RV_long_memory}
\bibliographystyle{chicago}

\section*{Appendix \label{appWavelets}}
\footnotesize

\subsection*{The continuous wavelet transform}

The continuous wavelet transform (CWT) $W_x(j,s)$ is obtained by projecting a specific wavelet $\psi(.)$ onto the examined time series $x_t\in L^2(\mathbb{R})$, i.e.,
\begin{equation}
W_x (j,s)=\int_{-\infty}^\infty x_t\frac{1}{\sqrt j}\overline{\psi \left( \frac{t-s}{j}\right)} dt.
\label{eq7}
\end{equation}
A wavelet is a real-valued square integrable function, $\psi\in L^2(\mathbb{R})$\footnote{A function $x(t)$ is called a square integrable if $\int_{-\infty }^{\infty } x(t)^2 dt<\infty$.}, defined as:
\begin{equation}
\psi_{j,s}(t)=\frac{1}{\sqrt j}\psi \left( \frac{t-s}{j}\right),
\end{equation}
where the term $1/\sqrt{j}$ denotes a normalization factor. Parameter $s$ determines the exact position of the wavelet, whereas the scale parameter $j$ defines how the wavelet is stretched or dilated. Scale has an inverse relation to frequency; thus lower (higher) scale means a more (less) compressed wavelet, which is able to detect higher (lower) frequencies of a time series. A wavelet needs to satisfy admissibility condition, $C_{\psi}=\int_{0}^{\infty }\frac{\mid\Psi(f)\mid^2}{f}df<\infty$ where $\Psi(f)$ is the Fourier transform of a wavelet $\psi(.)$. Further, the wavelet is usually normalized to have unit energy, i.e., $\int_{-\infty }^{\infty }\psi^2 (t)dt=1$. In the analysis of wavelet coherence we use the Morlet wavelet, defined as: $\psi^M(t)=\frac{1}{\pi^{1/4}}e^{i\omega_0 t}e^{-t^2/2}$, where $\omega_0$ denotes the central frequency of the wavelet.  We set $\omega_0=6$, which is often used in economic applications -- see for example \cite{Conraria2008, RuaNunes2009}. The Morlet wavelet belongs to the family of complex or analytic wavelets, hence this wavelet has both real and imaginary parts, allowing us to study both amplitude and phase.

An important feature of the CWT is the ability to decompose and then subsequently perfectly reconstruct a time series $x_t\in L^2(\mathbb{R})$:
\begin{equation}
x_t=\frac{1}{C_\psi}\int_0^\infty \left[\int_{-\infty}^\infty  W_x (j,s) \psi_{j,s}(t) ds \right] \frac{dj}{j^2} ,\hspace{5 mm}s>0.
\end{equation}
Moreover, the CWT preserves the energy of the examined time series,  
\begin{equation}
\| x \|^2 =\frac{1}{C_\psi}\int_0^\infty  \left[\int_{-\infty}^\infty  \left| W_x (j,s)\right|^2 ds \right] \frac{dj}{j^2}.
\end{equation}

\subsubsection*{The maximal overlap discrete wavelet transform (MODWT)}
Since we work with real data, we need some form of sampling to compute the estimators, i.e., we have to use a suitable form of discretization. In this work, we use the maximal overlap discrete wavelet transformation (MODWT), a translation-invariant type of discrete wavelet transformation. Furthermore, the MODWT does not use a downsampling procedure as in the case of the discrete wavelet transform\footnote{For a definition and detailed discussion of the discrete wavelet transform see \cite{Mallat98}, \cite{PercivalWalden2000} and \cite{Gencay2002}} (DWT), so the wavelet and scaling coefficient vectors at all scales have equal length. As a consequence, the MODWT is not restricted to sample sizes that are powers of two. This feature is very important for the analysis of real market data, since this limitation of DWT is usually too restrictive. For more details about the MODWT see \cite{Mallat98}, \cite{PercivalWalden2000} and \cite{Gencay2002}.

The MODWT is a very convenient tool for variance and energy analysis of a time series in the time-frequency domain. \cite{Percival1995} demonstrates the advantages of the MODWT estimator of variance over the DWT estimator. \cite{Serroukh2000} analyze the statistical properties of the MODWT variance estimator for (locally) non-stationary and non-Gaussian processes.

\subsubsection*{Definition of MODWT filters}

First, let us introduce the MODWT scaling and wavelet filters $\tilde{g}_{l}$ and $\tilde{h}_{l}$, $l=0,1,\ldots,L-1$, where $L$ denotes the length of the wavelet filter. For example, the Daubechies D(4) wavelet filter has length $L=4$ \citep{Daubechies1992}. Generally, the scaling filter is a low-pass filter whereas the wavelet filter is a high-pass filter. There are three basic properties that both the MODWT filters must fulfill. Let us show these properties for the MODWT wavelet filter:
\begin{equation}
\sum_{l=0}^{L-1}\tilde{h}_{l}=0,\hspace{2mm}  \sum_{l=0}^{L-1}\tilde{h}_{l}^2=1/2, \hspace{2mm}  \sum_{l=-\infty}^{\infty} \tilde{h}_{l} \tilde{h}_{l+2n}=0, \hspace{2mm} n\in\ZN,
\end{equation}
and for the MODWT scaling filter:
\begin{equation}
\sum_{l=0}^{L-1}\tilde{g}_{l}=1,\hspace{2mm}  \sum_{l=0}^{L-1}\tilde{g}_{l}^2=1/2, \hspace{2mm}  \sum_{l=-\infty}^{\infty} \tilde{g}_{l} \tilde{g}_{l+2n}=0, \hspace{2mm} n\in\ZN.
\end{equation}
The transfer function of a MODWT filter $\{\tilde{h}_{l}\}$ at frequency $f$ is defined via the Fourier transform as:
\begin{equation}
\tilde{H(f)}\equiv \sum_{l=-\infty}^{\infty}\tilde{h}_{l} e^{-i2\pi f l}=\sum_{l=0}^{L-1}\tilde{h}_{l} e^{-i2\pi f l},
\end{equation}
with the squared gain function defined as: $\tilde{\mathcal{H}}(f)\equiv\vert \tilde{H(f)} \vert ^2.$

\subsubsection*{Pyramid algorithm}

We obtain the MODWT wavelet and scaling coefficients using the pyramid algorithm. The wavelet coefficients at the first scale ($j=1$) are obtained via filtering $x_t$ with the MODWT wavelet and scaling filters \citep{PercivalWalden2000}: 
\begin{equation}
\widetilde{W}_{1,s}\equiv\sum_{l=0}^{L-1}\tilde{h}_{l}x_{s-l mod N}, \hspace{5mm} \widetilde{V}_{1,s}\equiv\sum_{l=0}^{L-1}\tilde{g}_{l}x_{s-l mod N}.
\end{equation}
For the second stage of the algorithm, we replace $x_t$ with the scaling coefficients $\widetilde{V}_{1,s}$ and after the filtering we obtain wavelet coefficients at scale $j=2$ as:
\begin{equation}
\widetilde{W}_{2,s}\equiv\sum_{l=0}^{L-1}\tilde{h}_{l}\widetilde{V}_{1,s-l mod N}, \hspace{5mm} \widetilde{V}_{2,s}\equiv\sum_{l=0}^{L-1}\tilde{g}_{l}\widetilde{V}_{1,s-l mod N}.
\end{equation}
After two stages of the pyramid algorithm we have two vectors of the MODWT wavelet coefficients $ \mathbf{\widetilde{W}}_1, \mathbf{\widetilde{W}}_{2}$ and one vector of the MODWT wavelet scaling coefficients at scale two $\mathbf{\widetilde{V}}_2$. Vector $\mathbf{\widetilde{W}}_1$ represents wavelet coefficients at the frequency band $f\in[1/4,1/2]$, $\mathbf{\widetilde{W}}_2$: $f\in[1/8,1/4]$ and $\mathbf{\widetilde{V}}_2$: $f\in[0,1/8]$. The $j$-th level MODWT coefficients are in the form: 
\begin{equation}
\widetilde{W}_{j,s}\equiv\sum_{l=0}^{L-1}\tilde{h}_{l}\widetilde{V}_{j-1,s-l mod N}, \hspace{5mm} \widetilde{V}_{j,s}\equiv\sum_{l=0}^{L-1}\tilde{g}_{l}\widetilde{V}_{j-1,s-l mod N}, \hspace{3mm} j=1,2,\ldots, J.
\end{equation}
where $J\le log_2(T)$. Generally, the $j$-th level wavelet coefficients in the vector $\mathbf{\widetilde{W}}_j$ represents frequency bands $f\in[1/2^{j+1},1/2^{j}]$ wheres the $j$-th level scaling coefficients in the vector $\mathbf{\widetilde{V}}_j$ represents $f\in[0,1/2^{j+1}]$.

\subsubsection*{Energy decomposition of a stochastic process}

For our analysis, it is important to show that we are able to decompose the energy of a stochastic process on a scale-by-scale basis, i.e., we can get the energy contribution of every level $j$, with the maximum level of decomposition $J\le\log_2 T$. 

\begin{prop} Energy decomposition in discrete time \\
\label{propenergydecomposition}
The energy of the time series $x_t$, $t=1,\ldots,T-1$ can be decomposed on a scale-by-scale basis $J\le\log_2 N$ so that 
\begin{equation}
\label{energydec}
\|\mathbf{X}\|^2=\sum_{j=1}^{J} \|\widetilde{\mathbf{W}}_j\|^2 +\|\widetilde{\mathbf{V}}_{J}\|^2
\end{equation}
where $\|\mathbf{X} \| ^2=\sum_{t=0}^{T-1}x_t^2$,  $\|\widetilde{\mathbf{W}}_j \| ^2=\sum_{s=0}^{T-1}W_{j,s}^2$, $\|\widetilde{\mathbf{V}}_J \| ^2=\sum_{s=0}^{T-1}V_{J,s}^2\,$. $\widetilde{\mathbf{W}}_j$ and $\widetilde{\mathbf{V}}_j$ are $T$ dimensional vectors of the $j$-th level MODWT wavelet and scaling coefficients. 
\end{prop}
Proof of the energy decomposition using the MODWT can be found in \cite{PercivalWalden2000} and \cite{barunik}.

%


\subsubsection*{Wavelet variance}

For a real-valued covariance stationary stochastic process $x_t$ $t=1,2,\ldots, T$ with mean zero, or a covariance stationary process after $d-$th backward differences\footnote{$L\ge2d$ where $L$ is the length of a wavelet filter. In our analysis we use Daubechies D(4) filter with $L=4$.}, the sequence of the MODWT wavelet coefficients $\widetilde{W}_x(j,s)$, for all $j,s>0$ ,  is also stationary process with mean zero \citep{PercivalWalden2000}. Variance of the wavelet coefficients at a scale $j$ is the wavelet variance, i.e., 

\begin{equation}
\nu_x^2(j)=var\left( \widetilde{W}_x(j,s)\right)
\end{equation}
While the variance of a covariance stationary process $x_t$ is equal to the integral of the spectral density function $S_x (.)$, then for the wavelet variance at a particular level $j$ the variance of a wavelet coefficients $\widetilde{W}_x(j,s)$, has the spectral density function $S_{(x)j}(.)$:
\begin{equation}
\nu_x^2(j) =\int_{-1/2}^{1/2}S_{(x)j}(f)df =\int_{-1/2}^{1/2} \widetilde{\mathcal{H}}_j (f)S_{x}(f)df, 
\end{equation}
where $\widetilde{\mathcal{H}}_j (f)$ is the squared gain function of a wavelet filter $\tilde{h}_{j}$, \citep{Gencay2002, PercivalWalden2000}. Since the variance of a process $x_t$ is a sum of the contribution of variances at all scales, for $J\rightarrow\infty$ we can write:
\begin{equation}
var(x_t)=\sum_{j=1}^\infty \nu_x^2(j) 
\end{equation}
However, for a finite number of scales we have:
\begin{equation}
var\left(x_t\right)=\int_{-1/2}^{1/2}S_{x}(f)df=\sum_{j=1}^J \nu_x^2(j) +var\left(\widetilde{V}_X(J,s)\right) \hspace{5mm} j=1,2,\ldots, J
\end{equation}
Estimator of a wavelet variance at the $j$-th scale is defined as:
\begin{equation}
\hat{\nu}_x^2(j)\equiv\frac{1}{M_j}\sum_{s=L_j-1}^{T-1}  \widetilde{W}_x(j,s)
\end{equation}
where $M_j=T-L_j+1>0$ is number of $j$-th level MODWT coefficients unaffected by a boundary conditions\footnote{for  detailed discussion about boundary conditions see \cite{PercivalWalden2000}.}. In case we take all MODWT wavelet coefficients $T$ we obtain the biased MODWT variance estimator. However as $T\rightarrow \infty$ then the ratio $\frac{T}{M_j}$ goes to unity, so consequently the estimators gives more or less the same results. \cite{Serroukh2000} derived the asymptotic distribution of the MODWT wavelet variance estimator for Gaussian, non-Gaussian and non-linear processes.

\subsection*{Wavelet covariance}

Let $x_t$ and $y_t$ be covariance stationary processes with the square integrable spectral density functions $S_x (.)$, $S_y(.)$ and cross spectra $S_{xy} (.)$. Since we use the Daubechies family wavelet with the length $L=4$, we can use generally non-stationary process, that is stationary after $d$-th difference, where $d\le L/2$. 
The wavelet covariance of $x_t$ and $y_t$ at level $j$ is then defined as:

\begin{equation}
\gamma_{xy}(j)=Cov\left( \widetilde{W}_x(j,s),\, \widetilde{W}_y(j,s)\right)
\end{equation}

For a particular level of decomposition $J\le log_2(T)$, the covariance of $x_t$ and $y_t$ is a sum of covariances of the MODWT wavelet coefficients $\gamma_{xy}(j)$ at all scales $j=1,2,\ldots,J$ and covariance of the scaling coefficients $\widetilde{V}_x(J,s)$ a scale $J$:
\begin{equation}
\label{WaveCov}
Cov\left(x_t,y_t\right)=Cov\left(\widetilde{V}_x(J,s),\, \widetilde{V}_y(J,s)\right)+\sum_{j=1}^J \gamma_{xy}(j)
\end{equation}
In a case when $J\rightarrow \infty$, the covariance of scaling coefficients $\left(\widetilde{V}_x(J,s),\, \widetilde{V}_y(J,s)\right)$ goes to zero \citep{WGPTech1999}, hence we can rewrite \ref{WaveCov} as:
\begin{equation}
\label{WaveCov2eq}
Cov\left(x_t,y_t\right)=\sum_{j=1}^\infty \gamma_{xy}(j)
\end{equation}

For processes  $x_t$ and $y_t$ defined above, the estimator of a wavelet covariance at a level $j$ is defined as 
\begin{equation}
\hat{\gamma}_{xy}(j)=\frac{1}{M_j}\sum_{t=L_j-1}^{N-1} \widetilde{W}_x(j,s) \widetilde{W}_y(j,s),
\end{equation}
where $M_j=T-L_j+1>0$ is number of $j$-th level MODWT coefficients for both processes that are unaffected by boundary conditions. \cite{WGPTech1999} prove that for the Gaussian processes $x_t$ and $y_t$, the MODWT  estimator of wavelet covariance is unbiased and asymptotically normally distributed.



\newpage

\begin{table}[h]
\caption{Long memory estimates of CIV1, CIV2, MFIV measures of implied volatility, RV and JWTSRV measures of realized volatility series. Long memory parameter $d$ is estimated using GPH estimator with different frequency bands. Standard errors are provided in the parentheses.}
\begin{center}
\footnotesize
\begin{tabular}{lrrrrrrrrrrrrrr}
\toprule 
& \multicolumn{2}{c}{CIV1} & & \multicolumn{2}{c}{CIV2} & & \multicolumn{2}{c}{MFIV} & & \multicolumn{2}{c}{RV} & & \multicolumn{2}{c}{JWTSRV} \\
 \cmidrule{2-3} \cmidrule{5-6} \cmidrule{8-9} \cmidrule{11-12}  \cmidrule{14-15} 
 \multicolumn{3}{l}{\textbf{SPX monthly }}\\
 &\\
$[T^{0.6}]$ & 0.581 & (0.144) & &0.573 & (0.144) & &0.582 & (0.144) & &0.515 & (0.144) & &0.534 & (0.144)\\
$[T^{0.7}]$ & 0.523 & (0.115) & &0.517 & (0.115) & &0.546 & (0.115) & &0.538 & (0.115) & &0.555 & (0.115)\\
$[T^{0.8}]$ & 0.641 & (0.091) & &0.636 & (0.091) & &0.665 & (0.091) & &0.776 & (0.091) & &0.776 & (0.091)\\
&\\
 \multicolumn{3}{l}{\textbf{SPX bi-weekly}}\\
 &\\
$[T^{0.6}]$ & 0.654 & (0.144) & &0.662 & (0.144) & &0.494 & (0.144) & &0.526 & (0.144) & &0.520 & (0.144)\\
$[T^{0.7}]$ & 0.782 & (0.118) & &0.784 & (0.118) & &0.702 & (0.118) & &0.693 & (0.118) & &0.637 & (0.118)\\
$[T^{0.8}]$ & 0.711 & (0.094) & &0.683 & (0.094) & &0.527 & (0.094) & &0.779 & (0.094) & &0.752 & (0.094)\\
&\\
 \multicolumn{3}{l}{\textbf{DAX monthly}}\\
 &\\
$[T^{0.6}]$ & 0.626 & (0.139) & &0.631 & (0.139) & &0.570 & (0.139) & &0.448 & (0.139) & &0.486 & (0.139)\\
$[T^{0.7}]$ & 0.506 & (0.112) & &0.509 & (0.112) & &0.343 & (0.112) & &0.498 & (0.112) & &0.531 & (0.112)\\
$[T^{0.8}]$ & 0.574 & (0.091) & &0.577 & (0.091) & &0.300 & (0.091) & &0.613 & (0.091) & &0.633 & (0.091)\\
&\\
 \multicolumn{3}{l}{\textbf{DAX bi-weekly}}\\
 &\\
$[T^{0.6}]$ & 0.674 & (0.112) & &0.680 & (0.112) & &0.543 & (0.112) & &0.425 & (0.112) & &0.441 & (0.112)\\
$[T^{0.7}]$ & 0.572 & (0.087) & &0.573 & (0.087) & &0.371 & (0.087) & &0.544 & (0.087) & &0.542 & (0.087)\\
$[T^{0.8}]$ & 0.700 & (0.067) & &0.701 & (0.067) & &0.333 & (0.067) & &0.689 & (0.067) & &0.706 & (0.067)\\
&\\
  \bottomrule
\end{tabular}
\end{center}
\label{GPH_raw}
\end{table}%

\begin{table}[h]
\caption{Estimates of the realized-implied volatility relation using OLS, WBLS and FMNBLS on the S\&P 500 data using monthly maturities to calculate implied volatility. Implied volatility is measured by CIV1, CIV2 and MFIV, realized volatility is measured by RV and JWTSRV. Standard errors are in parentheses.}
\begin{center}
\footnotesize
\begin{tabular}{lrrrrrrrrrrr}
\toprule 
& \multicolumn{11}{c}{\textbf{RV}} \\   
 \midrule 
& \multicolumn{3}{c}{CIV1} & & \multicolumn{3}{c}{CIV2} & & \multicolumn{3}{c}{MFIV} \\   
 \cmidrule{2-4} \cmidrule{6-8} \cmidrule{10-12}  
 & $\alpha$ & $\beta$ & $d$ & & $\alpha$ & $\beta$ & $d$ & & $\alpha$ & $\beta$ & $d$ \\ 
 \cmidrule{2-4} \cmidrule{6-8} \cmidrule{10-12} 
OLS &0.005 & 0.684 & 0.376 &    &0.005 & 0.657 & 0.384 &    &0.006 & 0.609 & 0.342 \\ 
  &(0.005) & (0.068) & (0.177) &    &(0.005) & (0.066) & (0.177) &    &(0.005) & (0.060) & (0.177)  \\ 
 \cmidrule{2-4} \cmidrule{6-8} \cmidrule{10-12}  
WBLS &-0.000 & 0.822 & 0.058 &    &-0.000 & 0.795 & 0.064 &    &-0.000 & 0.736 & 0.041 \\ 
  &(0.000) & (0.027) & (0.115) &    &(0.000) & (0.026) & (0.115) &    &(0.000) & (0.024) & (0.115)  \\ 
 \cmidrule{1-4} \cmidrule{6-8} \cmidrule{10-12}  
FMNBLS & $\alpha$ & $\beta$ & $d$ & & $\alpha$ & $\beta$ & $d$ & & $\alpha$ & $\beta$ & $d$ \\ 
 \cmidrule{1-4} \cmidrule{6-8} \cmidrule{10-12}   
$[T^{0.4},T^{0.6}]$ &-0.009 & 0.908 & 0.070 &    &-0.009 & 0.891 & 0.084 &    &-0.007 & 0.798 & 0.045 \\ 
  &   & (0.038) & (0.144) &    &   & (0.025) & (0.144) &    &   &    & (0.144) \\ 
$[T^{0.4},T^{0.7}]$ &-0.010 & 0.926 & 0.056 &    &-0.010 & 0.904 & 0.062 &    &-0.007 & 0.807 & 0.037 \\ 
  &   & (0.048) & (0.115) &    &   & (0.055) & (0.115) &    &   & (0.026) & (0.115) \\ 
$[T^{0.4},T^{0.8}]$ &-0.009 & 0.917 & 0.109 &    &-0.009 & 0.893 & 0.111 &    &-0.007 & 0.797 & 0.091 \\ 
  &   & (0.097) & (0.091) &    &   & (0.100) & (0.091) &    &   &    & (0.091) \\ 
$[T^{0.5},T^{0.6}]$ &-0.012 & 0.962 & 0.088 &    &-0.013 & 0.944 & 0.102 &    &-0.009 & 0.832 & 0.055 \\ 
  &   & (0.020) & (0.144) &    &   & (0.044) & (0.144) &    &   & (0.047) & (0.144) \\ 
$[T^{0.5},T^{0.7}]$ &-0.013 & 0.986 & 0.071 &    &-0.014 & 0.962 & 0.077 &    &-0.010 & 0.844 & 0.046 \\ 
  &   & (0.060) & (0.115) &    &   & (0.067) & (0.115) &    &   & (0.001) & (0.115) \\ 
$[T^{0.5},T^{0.8}]$ &-0.012 & 0.969 & 0.116 &    &-0.013 & 0.943 & 0.119 &    &-0.009 & 0.827 & 0.093 \\ 
 &   & (0.029) & (0.091) &    &   & (0.036) & (0.091) &    &   &    & (0.091) \\ 
 \midrule 
& \multicolumn{11}{c}{\textbf{JWTSRV}} \\   
 \midrule 
& \multicolumn{3}{c}{CIV1} & & \multicolumn{3}{c}{CIV2} & & \multicolumn{3}{c}{MFIV} \\   
 \cmidrule{2-4} \cmidrule{6-8} \cmidrule{10-12}  
 & $\alpha$ & $\beta$ & $d$ & & $\alpha$ & $\beta$ & $d$ & & $\alpha$ & $\beta$ & $d$ \\ 
 \cmidrule{2-4} \cmidrule{6-8} \cmidrule{10-12} 
OLS &0.004 & 0.697 & 0.424 &    &0.004 & 0.669 & 0.431 &    &0.005 & 0.621 & 0.393 \\ 
  &(0.005) & (0.067) & (0.177) &    &(0.005) & (0.065) & (0.177) &    &(0.005) & (0.060) & (0.177)  \\ 
 \cmidrule{2-4} \cmidrule{6-8} \cmidrule{10-12}  
WBLS &-0.000 & 0.842 & 0.069 &    &-0.000 & 0.814 & 0.074 &    &0.000 & 0.753 & 0.054 \\ 
  &(0.000) & (0.027) & (0.115) &    &(0.000) & (0.027) & (0.115) &    &(0.000) & (0.024) & (0.115)  \\ 
 \cmidrule{1-4} \cmidrule{6-8} \cmidrule{10-12}  
FMNBLS & $\alpha$ & $\beta$ & $d$ & & $\alpha$ & $\beta$ & $d$ & & $\alpha$ & $\beta$ & $d$ \\ 
 \cmidrule{1-4} \cmidrule{6-8} \cmidrule{10-12}   
$[T^{0.4},T^{0.6}]$ &-0.011 & 0.946 & 0.099 &    &-0.012 & 0.928 & 0.113 &    &-0.009 & 0.831 & 0.076 \\ 
  &   & (0.035) & (0.144) &    &   & (0.052) & (0.144) &    &   & (0.038) & (0.144) \\ 
$[T^{0.4},T^{0.7}]$ &-0.012 & 0.956 & 0.068 &    &-0.012 & 0.934 & 0.074 &    &-0.009 & 0.833 & 0.051 \\ 
  &   & (0.057) & (0.115) &    &   & (0.064) & (0.115) &    &   & (0.016) & (0.115) \\ 
$[T^{0.4},T^{0.8}]$ &-0.011 & 0.936 & 0.106 &    &-0.011 & 0.912 & 0.108 &    &-0.008 & 0.815 & 0.088 \\ 
  &   & (0.106) & (0.091) &    &   & (0.110) & (0.091) &    &   &    & (0.091) \\ 
$[T^{0.5},T^{0.6}]$ &-0.015 & 1.006 & 0.120 &    &-0.016 & 0.987 & 0.135 &    &-0.012 & 0.869 & 0.089 \\ 
  &   & (0.054) & (0.144) &    &   & (0.069) & (0.144) &    &   & (0.020) & (0.144) \\ 
$[T^{0.5},T^{0.7}]$ &-0.015 & 1.020 & 0.085 &    &-0.016 & 0.996 & 0.090 &    &-0.012 & 0.874 & 0.062 \\ 
  &   & (0.071) & (0.115) &    &   & (0.076) & (0.115) &    &   & (0.029) & (0.115) \\ 
$[T^{0.5},T^{0.8}]$ &-0.014 & 0.992 & 0.113 &    &-0.014 & 0.965 & 0.116 &    &-0.010 & 0.847 & 0.091 \\ 
 &   & (0.025) & (0.091) &    &   & (0.033) & (0.091) &    &   &    & (0.091) \\ 
  \bottomrule
\end{tabular}
\end{center}
\label{SPX29}
\end{table}%

\begin{table}[h]
\caption{Estimates of the realized-implied volatility relation using OLS, WBLS and FMNBLS on the S\&P 500 data using bi-weekly maturities to calculate implied volatility. Implied volatility is measured by CIV1, CIV2 and MFIV, realized volatility is measured by RV and JWTSRV. Standard errors are in parentheses.}
\begin{center}
\footnotesize
\begin{tabular}{lrrrrrrrrrrr}
\toprule 
& \multicolumn{11}{c}{\textbf{RV}} \\   
 \midrule 
& \multicolumn{3}{c}{CIV1} & & \multicolumn{3}{c}{CIV2} & & \multicolumn{3}{c}{MFIV} \\   
 \cmidrule{2-4} \cmidrule{6-8} \cmidrule{10-12}  
 & $\alpha$ & $\beta$ & $d$ & & $\alpha$ & $\beta$ & $d$ & & $\alpha$ & $\beta$ & $d$ \\ 
 \cmidrule{2-4} \cmidrule{6-8} \cmidrule{10-12} 
OLS &-0.007 & 0.938 & 0.161 &    &-0.006 & 0.883 & 0.185 &    &-0.003 & 0.631 & 0.343 \\ 
  &(0.004) & (0.082) & (0.177) &    &(0.004) & (0.079) & (0.177) &    &(0.004) & (0.068) & (0.177)  \\ 
 \cmidrule{2-4} \cmidrule{6-8} \cmidrule{10-12}  
WBLS &0.000 & 1.001 & 0.316 &    &0.000 & 0.945 & 0.329 &    &0.000 & 0.752 & 0.390 \\ 
  &(0.000) & (0.041) & (0.118) &    &(0.000) & (0.039) & (0.118) &    &(0.001) & (0.042) & (0.118)  \\ 
 \cmidrule{1-4} \cmidrule{6-8} \cmidrule{10-12}  
FMNBLS & $\alpha$ & $\beta$ & $d$ & & $\alpha$ & $\beta$ & $d$ & & $\alpha$ & $\beta$ & $d$ \\ 
 \cmidrule{1-4} \cmidrule{6-8} \cmidrule{10-12}   
$[T^{0.4},T^{0.6}]$ &-0.012 & 1.058 & 0.200 &    &-0.012 & 1.023 & 0.218 &    &-0.015 & 0.839 & 0.200 \\ 
  &   & (0.068) & (0.144) &    &   & (0.072) & (0.144) &    &   & (0.127) & (0.144) \\ 
$[T^{0.4},T^{0.7}]$ &-0.013 & 1.063 & 0.316 &    &-0.013 & 1.029 & 0.329 &    &-0.012 & 0.797 & 0.360 \\ 
  &   & (0.058) & (0.118) &    &   & (0.063) & (0.118) &    &   & (0.100) & (0.118) \\ 
$[T^{0.4},T^{0.8}]$ &-0.014 & 1.107 & 0.380 &    &-0.016 & 1.095 & 0.388 &    &-0.015 & 0.853 & 0.279 \\ 
  &   & (0.151) & (0.094) &    &   & (0.166) & (0.094) &    &   & (0.149) & (0.094) \\ 
$[T^{0.5},T^{0.6}]$ &-0.016 & 1.142 & 0.220 &    &-0.016 & 1.105 & 0.239 &    &-0.024 & 0.993 & 0.218 \\ 
  &   & (0.084) & (0.144) &    &   & (0.086) & (0.144) &    &   & (0.143) & (0.144) \\ 
$[T^{0.5},T^{0.7}]$ &-0.016 & 1.146 & 0.325 &    &-0.016 & 1.110 & 0.341 &    &-0.022 & 0.960 & 0.363 \\ 
  &   & (0.067) & (0.118) &    &   & (0.073) & (0.118) &    &   & (0.110) & (0.118) \\ 
$[T^{0.5},T^{0.8}]$ &-0.019 & 1.208 & 0.380 &    &-0.020 & 1.193 & 0.390 &    &-0.025 & 1.021 & 0.265 \\ 
 &   & (0.147) & (0.094) &    &   & (0.162) & (0.094) &    &   & (0.153) & (0.094) \\ 
 \midrule 
& \multicolumn{11}{c}{\textbf{JWTSRV}} \\   
 \midrule 
& \multicolumn{3}{c}{CIV1} & & \multicolumn{3}{c}{CIV2} & & \multicolumn{3}{c}{MFIV} \\   
 \cmidrule{2-4} \cmidrule{6-8} \cmidrule{10-12}  
 & $\alpha$ & $\beta$ & $d$ & & $\alpha$ & $\beta$ & $d$ & & $\alpha$ & $\beta$ & $d$ \\ 
 \cmidrule{2-4} \cmidrule{6-8} \cmidrule{10-12} 
OLS &-0.007 & 0.939 & 0.176 &    &-0.006 & 0.883 & 0.192 &    &-0.003 & 0.634 & 0.363 \\ 
  &(0.004) & (0.083) & (0.177) &    &(0.004) & (0.080) & (0.177) &    &(0.005) & (0.069) & (0.177)  \\ 
 \cmidrule{2-4} \cmidrule{6-8} \cmidrule{10-12}  
WBLS &-0.000 & 1.001 & 0.258 &    &-0.000 & 0.945 & 0.267 &    &0.000 & 0.750 & 0.337 \\ 
  &(0.000) & (0.040) & (0.118) &    &(0.000) & (0.039) & (0.118) &    &(0.001) & (0.042) & (0.118)  \\ 
 \cmidrule{1-4} \cmidrule{6-8} \cmidrule{10-12}  
FMNBLS & $\alpha$ & $\beta$ & $d$ & & $\alpha$ & $\beta$ & $d$ & & $\alpha$ & $\beta$ & $d$ \\ 
 \cmidrule{1-4} \cmidrule{6-8} \cmidrule{10-12}   
$[T^{0.4},T^{0.6}]$ &-0.012 & 1.052 & 0.206 &    &-0.012 & 1.019 & 0.226 &    &-0.014 & 0.829 & 0.224 \\ 
  &   & (0.074) & (0.144) &    &   & (0.079) & (0.144) &    &   & (0.135) & (0.144) \\ 
$[T^{0.4},T^{0.7}]$ &-0.012 & 1.043 & 0.258 &    &-0.012 & 1.006 & 0.269 &    &-0.011 & 0.774 & 0.317 \\ 
  &   & (0.127) & (0.118) &    &   & (0.153) & (0.118) &    &   & (0.079) & (0.118) \\ 
$[T^{0.4},T^{0.8}]$ &-0.014 & 1.088 & 0.342 &    &-0.015 & 1.073 & 0.351 &    &-0.015 & 0.834 & 0.277 \\ 
  &   & (0.130) & (0.094) &    &   & (0.147) & (0.094) &    &   & (0.145) & (0.094) \\ 
$[T^{0.5},T^{0.6}]$ &-0.017 & 1.152 & 0.232 &    &-0.017 & 1.113 & 0.252 &    &-0.024 & 0.990 & 0.240 \\ 
  &   & (0.094) & (0.144) &    &   & (0.097) & (0.144) &    &   & (0.150) & (0.144) \\ 
$[T^{0.5},T^{0.7}]$ &-0.016 & 1.133 & 0.271 &    &-0.016 & 1.091 & 0.281 &    &-0.020 & 0.929 & 0.321 \\ 
  &   & (0.047) & (0.118) &    &   & (0.054) & (0.118) &    &   & (0.088) & (0.118) \\ 
$[T^{0.5},T^{0.8}]$ &-0.019 & 1.204 & 0.345 &    &-0.020 & 1.183 & 0.356 &    &-0.025 & 1.005 & 0.262 \\ 
 &   & (0.131) & (0.094) &    &   & (0.147) & (0.094) &    &   & (0.149) & (0.094) \\ 
   \bottomrule
\end{tabular}
\end{center}
\label{SPX15}
\end{table}%

\begin{table}[h]
\caption{Estimates of the realized-implied volatility relation using OLS, WBLS and FMNBLS on the DAX data using monthly maturities to calculate implied volatility. Implied volatility is measured by CIV1, CIV2 and MFIV, realized volatility is measured by RV and JWTSRV. Standard errors are in parentheses.}
\begin{center}
\footnotesize
\begin{tabular}{lrrrrrrrrrrr}
\toprule 
& \multicolumn{11}{c}{\textbf{RV}} \\   
 \midrule 
& \multicolumn{3}{c}{CIV1} & & \multicolumn{3}{c}{CIV2} & & \multicolumn{3}{c}{MFIV} \\   
 \cmidrule{2-4} \cmidrule{6-8} \cmidrule{10-12}  
 & $\alpha$ & $\beta$ & $d$ & & $\alpha$ & $\beta$ & $d$ & & $\alpha$ & $\beta$ & $d$ \\ 
 \cmidrule{2-4} \cmidrule{6-8} \cmidrule{10-12} 
OLS &0.005 & 0.774 & -0.022 &    &0.005 & 0.748 & -0.032 &    &0.025 & 0.333 & 0.338 \\ 
  &(0.005) & (0.073) & (0.177) &    &(0.005) & (0.070) & (0.177) &    &(0.005) & (0.053) & (0.177)  \\ 
 \cmidrule{2-4} \cmidrule{6-8} \cmidrule{10-12}  
WBLS &-0.000 & 0.901 & -0.133 &    &0.000 & 0.869 & -0.130 &    &0.000 & 0.579 & -0.041 \\ 
  &(0.000) & (0.026) & (0.112) &    &(0.000) & (0.025) & (0.112) &    &(0.000) & (0.023) & (0.112)  \\ 
 \cmidrule{1-4} \cmidrule{6-8} \cmidrule{10-12}  
FMNBLS & $\alpha$ & $\beta$ & $d$ & & $\alpha$ & $\beta$ & $d$ & & $\alpha$ & $\beta$ & $d$ \\ 
 \cmidrule{1-4} \cmidrule{6-8} \cmidrule{10-12}   
$[T^{0.4},T^{0.6}]$ &0.001 & 0.835 & -0.251 &    &0.001 & 0.806 & -0.258 &    &-0.003 & 0.692 & -0.236 \\ 
  &   & (0.043) & (0.139) &    &   & (0.043) & (0.139) &    &   & (0.014) & (0.139) \\ 
$[T^{0.4},T^{0.7}]$ &0.001 & 0.844 & -0.121 &    &0.001 & 0.814 & -0.118 &    &-0.007 & 0.742 & -0.186 \\ 
  &   & (0.003) & (0.112) &    &   & (0.005) & (0.112) &    &   & (0.017) & (0.112) \\ 
$[T^{0.4},T^{0.8}]$ &-0.001 & 0.868 & -0.030 &    &-0.001 & 0.836 & -0.025 &    &-0.010 & 0.780 & -0.110 \\ 
  &   &    & (0.091) &    &   &    & (0.091) &    &   & (0.066) & (0.091) \\ 
$[T^{0.5},T^{0.6}]$ &0.004 & 0.789 & -0.204 &    &0.004 & 0.759 & -0.206 &    &-0.002 & 0.677 & -0.254 \\ 
  &   & (0.052) & (0.139) &    &   & (0.052) & (0.139) &    &   & (0.023) & (0.139) \\ 
$[T^{0.5},T^{0.7}]$ &0.002 & 0.825 & -0.108 &    &0.002 & 0.792 & -0.103 &    &-0.007 & 0.742 & -0.185 \\ 
  &   & (0.003) & (0.112) &    &   & (0.005) & (0.112) &    &   & (0.030) & (0.112) \\ 
$[T^{0.5},T^{0.8}]$ &-0.000 & 0.865 & -0.029 &    &-0.000 & 0.831 & -0.023 &    &-0.009 & 0.766 & -0.117 \\ 
 &   &    & (0.091) &    &   &    & (0.091) &    &   & (0.059) & (0.091) \\ 
 \midrule 
& \multicolumn{11}{c}{\textbf{JWTSRV}} \\   
 \midrule 
& \multicolumn{3}{c}{CIV1} & & \multicolumn{3}{c}{CIV2} & & \multicolumn{3}{c}{MFIV} \\   
 \cmidrule{2-4} \cmidrule{6-8} \cmidrule{10-12}  
 & $\alpha$ & $\beta$ & $d$ & & $\alpha$ & $\beta$ & $d$ & & $\alpha$ & $\beta$ & $d$ \\ 
 \cmidrule{2-4} \cmidrule{6-8} \cmidrule{10-12} 
OLS &0.003 & 0.764 & 0.028 &    &0.003 & 0.739 & 0.019 &    &0.022 & 0.331 & 0.385 \\ 
  &(0.005) & (0.070) & (0.177) &    &(0.005) & (0.067) & (0.177) &    &(0.005) & (0.051) & (0.177)  \\ 
 \cmidrule{2-4} \cmidrule{6-8} \cmidrule{10-12}  
WBLS &-0.000 & 0.909 & -0.132 &    &0.000 & 0.876 & -0.128 &    &0.000 & 0.585 & -0.033 \\ 
  &(0.000) & (0.024) & (0.112) &    &(0.000) & (0.023) & (0.112) &    &(0.000) & (0.022) & (0.112)  \\ 
 \cmidrule{1-4} \cmidrule{6-8} \cmidrule{10-12}  
FMNBLS & $\alpha$ & $\beta$ & $d$ & & $\alpha$ & $\beta$ & $d$ & & $\alpha$ & $\beta$ & $d$ \\ 
 \cmidrule{1-4} \cmidrule{6-8} \cmidrule{10-12}   
$[T^{0.4},T^{0.6}]$ &-0.001 & 0.830 & -0.237 &    &-0.001 & 0.800 & -0.242 &    &-0.005 & 0.683 & -0.243 \\ 
  &   & (0.044) & (0.139) &    &   & (0.045) & (0.139) &    &   & (0.013) & (0.139) \\ 
$[T^{0.4},T^{0.7}]$ &-0.003 & 0.859 & -0.127 &    &-0.003 & 0.829 & -0.123 &    &-0.010 & 0.738 & -0.184 \\ 
  &   & (0.003) & (0.112) &    &   & (0.005) & (0.112) &    &   & (0.018) & (0.112) \\ 
$[T^{0.4},T^{0.8}]$ &-0.004 & 0.883 & -0.056 &    &-0.004 & 0.851 & -0.051 &    &-0.013 & 0.778 & -0.122 \\ 
  &   &    & (0.091) &    &   &    & (0.091) &    &   & (0.058) & (0.091) \\ 
$[T^{0.5},T^{0.6}]$ &0.002 & 0.785 & -0.184 &    &0.002 & 0.754 & -0.184 &    &-0.005 & 0.674 & -0.253 \\ 
  &   & (0.052) & (0.139) &    &   & (0.052) & (0.139) &    &   & (0.022) & (0.139) \\ 
$[T^{0.5},T^{0.7}]$ &-0.001 & 0.837 & -0.114 &    &-0.001 & 0.805 & -0.109 &    &-0.010 & 0.743 & -0.180 \\ 
  &   & (0.003) & (0.112) &    &   & (0.005) & (0.112) &    &   & (0.034) & (0.112) \\ 
$[T^{0.5},T^{0.8}]$ &-0.003 & 0.871 & -0.053 &    &-0.003 & 0.837 & -0.047 &    &-0.012 & 0.769 & -0.128 \\ 
 &   &    & (0.091) &    &   &    & (0.091) &    &   & (0.053) & (0.091) \\  
   \bottomrule
\end{tabular}
\end{center}
\label{DAX28}
\end{table}%

\begin{table}[h]
\caption{Estimates of the realized-implied volatility relation using OLS, WBLS and FMNBLS on the DAX data using bi-weekly maturities to calculate implied volatility. Implied volatility is measured by CIV1, CIV2 and MFIV, realized volatility is measured by RV and JWTSRV. Standard errors are in parentheses.}
\begin{center}
\footnotesize
\begin{tabular}{lrrrrrrrrrrr}
\toprule 
& \multicolumn{11}{c}{\textbf{RV}} \\   
 \midrule 
& \multicolumn{3}{c}{CIV1} & & \multicolumn{3}{c}{CIV2} & & \multicolumn{3}{c}{MFIV} \\   
 \cmidrule{2-4} \cmidrule{6-8} \cmidrule{10-12}  
 & $\alpha$ & $\beta$ & $d$ & & $\alpha$ & $\beta$ & $d$ & & $\alpha$ & $\beta$ & $d$ \\ 
 \cmidrule{2-4} \cmidrule{6-8} \cmidrule{10-12} 
OLS &-0.001 & 0.917 & 0.021 &    &-0.001 & 0.876 & 0.020 &    &0.019 & 0.350 & 0.104 \\ 
  &(0.003) & (0.058) & (0.144) &    &(0.003) & (0.056) & (0.144) &    &(0.003) & (0.045) & (0.144)  \\ 
 \cmidrule{2-4} \cmidrule{6-8} \cmidrule{10-12}  
WBLS &0.000 & 1.092 & 0.232 &    &-0.000 & 1.049 & 0.231 &    &-0.000 & 0.628 & 0.153 \\ 
  &(0.000) & (0.034) & (0.087) &    &(0.000) & (0.033) & (0.087) &    &(0.000) & (0.033) & (0.087)  \\ 
 \cmidrule{1-4} \cmidrule{6-8} \cmidrule{10-12}  
FMNBLS & $\alpha$ & $\beta$ & $d$ & & $\alpha$ & $\beta$ & $d$ & & $\alpha$ & $\beta$ & $d$ \\ 
 \cmidrule{1-4} \cmidrule{6-8} \cmidrule{10-12}   
$[T^{0.4},T^{0.6}]$ &-0.000 & 0.889 & 0.043 &    &0.000 & 0.853 & 0.046 &    &-0.000 & 0.727 & 0.075 \\ 
  &   & (0.116) & (0.112) &    &   & (0.112) & (0.112) &    &   & (0.031) & (0.112) \\ 
$[T^{0.4},T^{0.7}]$ &-0.004 & 0.994 & 0.202 &    &-0.004 & 0.955 & 0.202 &    &-0.006 & 0.843 & 0.084 \\ 
  &   & (0.108) & (0.087) &    &   & (0.105) & (0.087) &    &   & (0.126) & (0.087) \\ 
$[T^{0.4},T^{0.8}]$ &-0.007 & 1.058 & 0.313 &    &-0.007 & 1.021 & 0.316 &    &-0.011 & 0.925 & 0.104 \\ 
  &   & (0.101) & (0.067) &    &   & (0.099) & (0.067) &    &   & (0.170) & (0.067) \\ 
$[T^{0.5},T^{0.6}]$ &-0.003 & 0.957 & 0.072 &    &-0.003 & 0.915 & 0.073 &    &-0.003 & 0.783 & 0.117 \\ 
  &   & (0.126) & (0.112) &    &   & (0.121) & (0.112) &    &   & (0.051) & (0.112) \\ 
$[T^{0.5},T^{0.7}]$ &-0.007 & 1.063 & 0.212 &    &-0.007 & 1.019 & 0.211 &    &-0.009 & 0.894 & 0.097 \\ 
  &   & (0.108) & (0.087) &    &   & (0.104) & (0.087) &    &   & (0.124) & (0.087) \\ 
$[T^{0.5},T^{0.8}]$ &-0.008 & 1.087 & 0.312 &    &-0.008 & 1.045 & 0.315 &    &-0.012 & 0.959 & 0.108 \\ 
 &   & (0.096) & (0.067) &    &   & (0.094) & (0.067) &    &   & (0.154) & (0.067) \\ 
 \midrule 
& \multicolumn{11}{c}{\textbf{JWTSRV}} \\   
 \midrule 
& \multicolumn{3}{c}{CIV1} & & \multicolumn{3}{c}{CIV2} & & \multicolumn{3}{c}{MFIV} \\   
 \cmidrule{2-4} \cmidrule{6-8} \cmidrule{10-12}  
 & $\alpha$ & $\beta$ & $d$ & & $\alpha$ & $\beta$ & $d$ & & $\alpha$ & $\beta$ & $d$ \\ 
 \cmidrule{2-4} \cmidrule{6-8} \cmidrule{10-12} 
OLS &-0.003 & 0.925 & 0.078 &    &-0.003 & 0.885 & 0.077 &    &0.017 & 0.356 & 0.148 \\ 
  &(0.003) & (0.056) & (0.144) &    &(0.003) & (0.055) & (0.144) &    &(0.003) & (0.044) & (0.144)  \\ 
 \cmidrule{2-4} \cmidrule{6-8} \cmidrule{10-12}  
WBLS &0.000 & 1.108 & 0.217 &    &-0.000 & 1.065 & 0.216 &    &-0.000 & 0.643 & 0.150 \\ 
  &(0.000) & (0.032) & (0.087) &    &(0.000) & (0.031) & (0.087) &    &(0.000) & (0.032) & (0.087)  \\ 
 \cmidrule{1-4} \cmidrule{6-8} \cmidrule{10-12}  
FMNBLS & $\alpha$ & $\beta$ & $d$ & & $\alpha$ & $\beta$ & $d$ & & $\alpha$ & $\beta$ & $d$ \\ 
 \cmidrule{1-4} \cmidrule{6-8} \cmidrule{10-12}   
$[T^{0.4},T^{0.6}]$ &-0.003 & 0.915 & 0.055 &    &-0.003 & 0.879 & 0.058 &    &-0.003 & 0.748 & 0.082 \\ 
  &   & (0.107) & (0.112) &    &   & (0.104) & (0.112) &    &   & (0.035) & (0.112) \\ 
$[T^{0.4},T^{0.7}]$ &-0.006 & 0.999 & 0.190 &    &-0.006 & 0.960 & 0.190 &    &-0.009 & 0.854 & 0.076 \\ 
  &   & (0.100) & (0.087) &    &   & (0.096) & (0.087) &    &   & (0.122) & (0.087) \\ 
$[T^{0.4},T^{0.8}]$ &-0.009 & 1.056 & 0.309 &    &-0.009 & 1.018 & 0.312 &    &-0.012 & 0.926 & 0.092 \\ 
  &   & (0.096) & (0.067) &    &   & (0.095) & (0.067) &    &   & (0.161) & (0.067) \\ 
$[T^{0.5},T^{0.6}]$ &-0.006 & 0.983 & 0.083 &    &-0.005 & 0.940 & 0.085 &    &-0.006 & 0.800 & 0.121 \\ 
  &   & (0.119) & (0.112) &    &   & (0.115) & (0.112) &    &   & (0.053) & (0.112) \\ 
$[T^{0.5},T^{0.7}]$ &-0.010 & 1.075 & 0.201 &    &-0.009 & 1.029 & 0.201 &    &-0.011 & 0.904 & 0.090 \\ 
  &   & (0.102) & (0.087) &    &   & (0.098) & (0.087) &    &   & (0.121) & (0.087) \\ 
$[T^{0.5},T^{0.8}]$ &-0.010 & 1.095 & 0.307 &    &-0.010 & 1.052 & 0.311 &    &-0.014 & 0.962 & 0.097 \\ 
 &   & (0.091) & (0.067) &    &   & (0.090) & (0.067) &    &   & (0.148) & (0.067) \\
   \bottomrule
\end{tabular}
\end{center}
\label{DAX15}
\end{table}%

\end{document}